\documentclass[preprint,review,12pt]{elsarticle}

\usepackage{amssymb}
\usepackage{amsmath,bm}
\usepackage{colortbl,booktabs}
\usepackage{tabularx}
\usepackage{threeparttable}
\usepackage{multicol,multirow}
\usepackage{subfigure}
\usepackage [font=footnotesize,labelfont=bf]{caption}
\usepackage{rotating}
\usepackage{tikz}
\usepackage{hyperref}

\usepackage{color}

\journal{Elsevier}

\begin{document}

\begin{frontmatter}

	\title{On zero-order consistency residue and background pressure for the conservative SPH fluid dynamics}

	\author{Feng Wang\texorpdfstring{\fnref{eqnote}}{}}
	\ead{feng.wang.aer@tum.de}
	\author{Xiangyu Hu\texorpdfstring{\corref{mycorrespondingauthor}}{}}
	\cortext[mycorrespondingauthor]{Corresponding author.}
	\ead{xiangyu.hu@tum.de}
	\address{Department of Mechanical Engineering, Technical University of Munich\\
		85748 Garching, Germany}

	\begin{abstract}

		As one of the major challenges for the conservative smoothed particle hydrodynamics (SPH) method, the zero-order consistency issue, although thought to be mitigated by the particle regularization scheme, such as the transport velocity formulation, significantly damps the flow in a long channel for both laminar and turbulent simulations.
		Building on this finding, this paper not only thoroughly analyzes the damping reason in this pressure-driven channel flow, but also relates this problem with the excessive numerical dissipation in the gravity-driven free-surface flow.
		The common root cause of the non-physical numerical damping in the two typical flow scenarios, the zero-order gradient consistency residue, is exposed.
		The adverse influence of the background pressure on the residue for the two scenarios is revealed and discussed.
		To comprehensively understand the behavior of the residue and mitigate its potential adverse effects, we conduct both theoretical analysis and numerical experiments focusing on the key sensitive factors.
		For studying the residue-induced non-physical energy dissipation in the gravity-driven free-surface flow, the water depth and input dynamic pressure in the inviscid standing wave case are tested.
		To investigate the velocity loss in the pressure-driven channel flow, we examine the effects of the channel length, resolution, and outlet pressure.
		The state-of-the-art reverse kernel gradient correction technique is introduced for the two typical flows, and proved to be effective in reducing the residue effect, but we find its correction capability is fundamentally limited.
		Finally, the FDA nozzle, an engineering benchmark, is tested to demonstrate the residue influence in a complex geometry, highlighting the necessity of correction schemes in scenarios with unavoidable high background pressure.
		This paper, for the first time, well defines a key drawback of the conservative SPH method that was generally supposed to be the accuracy problem, and shed light on engineers.

	\end{abstract}

	\begin{keyword}
		Smoothed particle hydrodynamic(SPH)  \sep Zero-order consistency  \sep Internal flow \sep Free-surface flow \sep Turbulence
	\end{keyword}

\end{frontmatter}
%
%
\section{Introduction}
\label{sec-intro}
Originally developed for astrophysical applications\cite{gingold1977smoothed}, the smoothed particle hydrodynamics (SPH) method has since been widely adopted for simulating fluid flows.
Due to its Lagrangian mesh-less characteristic, the SPH method has been showing superiority in simulating free surface flow and the channel flows that involve complex geometries or fluid-structure interaction (FSI).
However, when it comes to the fluid engineering practice, the conservation and consistency of the SPH method have been criticized.
This is because prior researches \cite{sun2018multi, liu2010smoothed, khayyer2017enhancement, liu2003smoothed, colagrossi2009theoretical} have suggested that the SPH method inherently involves a trade-off between the conservation and zero-order consistency, rendering it difficult to fully satisfy the both criteria simultaneously.
To strictly keep the consistency, the pressure differencing formulation (PDF)\cite{sun2018multi,khayyer2017enhancement}, which is fundamentally not conservative, is necessary.
On the contrary, in order to strictly maintain the conservative property, which is the priority in this work, the antisymmetric formulation\cite{bonet1999variational,litvinov2015towards} is indispensable.

Although the strictly conservative SPH (hereafter referred to as the conservative SPH for simplicity) method may violate the zero-order consistency condition during particle approximation\cite{liu2010smoothed}, for general flow simulation, many successful cases\cite{khayyer2017enhancement,zhang2021sphinxsys,adami2013transport} have been reported with the aid of some correction schemes, such as the particle shifting technique (PST)\cite{khayyer2017enhancement} and transport velocity formulation (TVF)\cite{adami2013transport}.

However, we recently find that the conservative SPH method encounters unexpected numerical dissipation in some particular situations, especially for the two typical types of flow discussed above, i.e., channel and free-surface flows.

\begin{figure}[htb!]
	\centering
	\includegraphics[trim = 0cm 0cm 0cm 9.4cm, clip,width=1.0\textwidth]{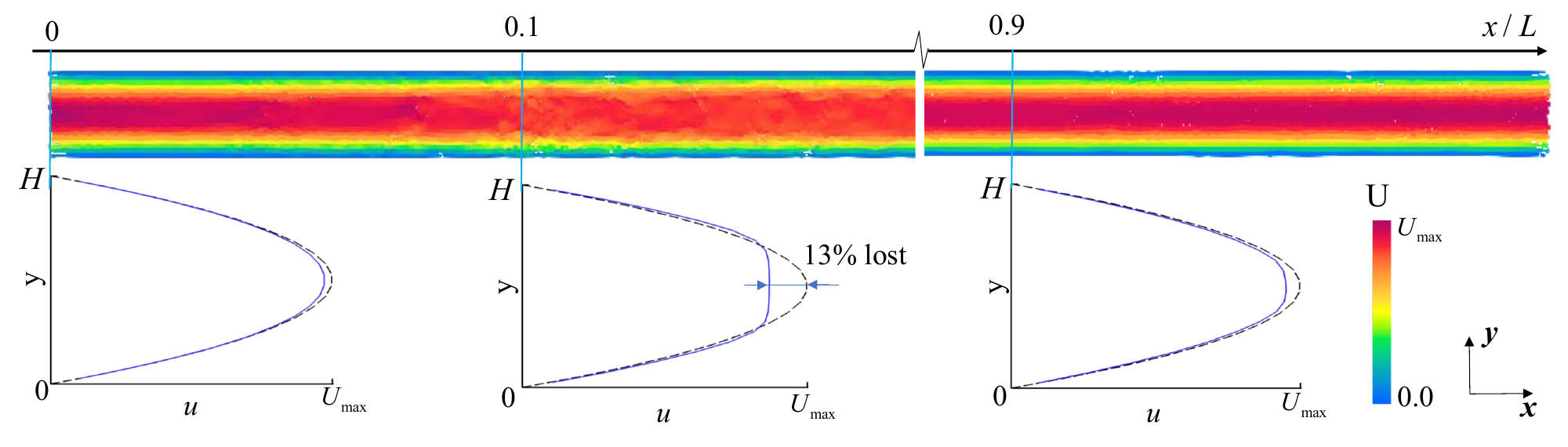}
	\caption
	{
		The velocity contour and cross-sectional velocity profiles in a long laminar channel($L/H=60$) computed by the conservative weakly compressible SPH method, where $L$ and $H$ is channel length and height, respectively, and $U$ is the axis velocity.
	}
	\label{fig-intro-damp}
\end{figure}

First, for channel flows, an apparent diffusion of the velocity field near the inlet region is observed when simulating the flow in a long channel with the velocity-inlet and pressure-outlet (VIPO) boundary conditions, as shown in Fig. \ref{fig-intro-damp}.
Meanwhile, the velocity profile at the cross-section becomes flattened, and the maximum velocity suffers an unusual dissipation, deviating from the expected parabolic shape of laminar flow.
While this problem has not, to the best of our knowledge, been reported in the literature, this may be attributed to the fact that most existing simulations focus on relatively short channels\cite{zhang2025dynamical} or adopt periodic boundary conditions\cite{sun2022coupled}.
Therefore, the reason for this problem still remains unclear.

Second, a similar unexpected numerical dissipation has been observed in the free surface flows for decades.
The dissipation leads to a significant under-prediction of the wave height\cite{wen2018improved} or amplitude\cite{liang2023study}, when using the conservative SPH method to simulate the wave propagation.
Given its critical importance in ocean engineering, this issue has been the subject of the extensive researches\cite{colagrossi2009theoretical,colagrossi2013smoothed,lyu2023derivation,zago2021overcoming, oger2007improved}.
However, although the existing studies find that the free-surface treatment\cite{colagrossi2009theoretical,colagrossi2013smoothed}, particle irregular distribution\cite{sun2017deltaplus}, enstrophy of the system\cite{lyu2023derivation}, smoothing length\cite{ren2023efficient,colagrossi2013smoothed}, and kernel type\cite{colagrossi2013smoothed} may be relevant with this issue, the underlying cause of the excessive numerical dissipation has not been conclusively identified.

In this work, we propose that the excessive numerical dissipation observed in the two representative types of flow stems from the previously mentioned zero-order consistency issue, and we define, for the first time to the best of our knowledge, the zero-order consistency residue to quantify this inconsistency.
More importantly, the inevitable background pressure, which is rarely discussed in the literature, is identified as a key contributing factor to the residual dissipation.
To support the hypothesis, theoretical and numerical analyses are conducted for the two types of flow, respectively.

For the free surface flow, the inviscid standing wave case, a well established benchmark in ocean engineering, is examined.
The effects of water depth and initial wave amplitude on residue-induced energy decay are analyzed.
For the channel flow, the laminar and turbulent flows in a velocity-inlet and pressure outlet (VIPO) straight channel are tested.
The influence of several factors, including channel length, resolution, outlet pressure, and the kernel gradient correction (KGC) scheme, on the residue effect is investigated.
Finally, the FDA (U.S. Food and Drug Administration) nozzle benchmark, a well validated test case in fluid engineering, is simulated to investigate the influence of complex geometry and to further underscore the pervasive and concerning nature of the zero-order consistency issue.

The remainder of this manuscript is organized as follows.
Section \ref{sec-numrical-methods} introduces the numerical methods used in this work, which are the preliminary works and not the part of our main contribution.
The concept of the zero-order consistency residue is introduced in Section \ref{sec-residue}.
Sections \ref{sec-freesurface} and \ref{section-channel-error} study the residue effect in the free-surface and channel flows, respectively.
The 2D and 3D FDA nozzle cases are tested and discussed in Section \ref{section-fda-channel}, and the concluding remarks are given in Section \ref{section-conclusion}.
The computational code of this work is released in the open-source SPHinXsys repository at https://github.com/Xiangyu-Hu/SPHinXsys.

\section{Preliminary: The WCSPH method for laminar and turbulent flows}
\label{sec-numrical-methods}
\subsection{Governing equations}
The conservation equations of mass and momentum for incompressible laminar and turbulent flows\cite{wang2025weakly} in the Lagrangian framework are
\begin{equation}
	\frac{\text{d} \rho}{\text{d} t} =  -\rho \nabla\cdot \mathbf v,
	\label{mass-equ}
\end{equation}
\begin{equation}
	\frac{\text{d} \mathbf v}{\text{d} t} =  - \frac{1}{\rho}\nabla p_{eff} +\nabla\cdot (\nu_{eff}\nabla \mathbf v) + \mathbf g,
	\label{momentum-equ}
\end{equation}
where $\mathbf{v}$ is the velocity, $\rho$ is the density, $\mathbf{g}$ is gravity,
$\frac{\text d}{\text d t}=\frac{\partial}{\partial t} + \mathbf v \cdot \nabla$ stands for material derivative.
$p_{eff}$ and $\nu_{eff}$ are the effective pressure and kinematic viscosity, respectively, and the expressions of the two variable differs in the laminar and turbulent simulations, as concluded in Table \ref{tab-unified-p-nu}, where $k$ refers to the turbulent kinetic energy, $\nu_l$ and $\nu_t$ are the kinematic molecular and eddy viscosity.
\begin{table}
	\centering
	\footnotesize
	\caption{Unified expressions for the effective pressure and kinematic viscosity.}
	\renewcommand{\arraystretch}{1.2}
	\begin{tabular}{l c c}
		\hline
		Viscous model    & $p_{eff}$                          & $\nu_{eff}$                 \\
		\hline
		Laminar          & $p$                                & $\nu$                       \\
		Turbulent (RANS) & $p_{eff} = p + \frac{2}{3} \rho k$ & $\nu_{eff} = \nu_l + \nu_t$ \\
		\hline
	\end{tabular}
	\label{tab-unified-p-nu}
\end{table}

To ensure the incompressibility, a stiff isothermal equation of state is used, as
\begin{equation}
	p = \rho_0 c^2_0 \left( \frac{\rho}{\rho_0} - 1\right),
	\label{eos}
\end{equation}
where $\rho_0$ is the reference density and $c_0$ refer to the sound speed.

For turbulent simulation, the two-equation $k - \epsilon$ RANS model is adopted and the transport equations are
\begin{equation}
	\frac{\text{d} k}{\text{d} t} = \bm{\tau_t} : \nabla \mathbf{v} -\epsilon+\nabla\cdot (D_k\nabla k),
	\label{k-equ}
\end{equation}
\begin{equation}
	\frac{\text{d} \epsilon}{\text{d} t} = C_1 \frac{\epsilon}{k} (\bm{\tau_t} : \nabla \mathbf{v} )-C_2 \frac{\epsilon^2}{k}+\nabla\cdot (D_\epsilon\nabla \epsilon),
	\label{epsilon-equ}
\end{equation}
where $D_k = \nu_l+\nu_t/\sigma_k$ and $D_\epsilon = \nu_l+\nu_t/\sigma_\epsilon$ are the diffusion coefficients for $k$ and $\epsilon$, respectively.
$\bm{\tau_t} =\nu_t(\nabla\mathbf v+\nabla\mathbf v^T)  - 2 k \mathbf I/3$ is the Reynolds stress tensor.
The kinematic eddy viscosity is calculated by $\nu_t= C_{\mu} k^2 / \epsilon $.
The empirical constants including $C_1$, $C_2$, $C_{\mu}$, $\sigma_k$ and $\sigma_\epsilon$ are the same from the original $k-\epsilon$ version\cite{launder1983numerical} and are concluded in Table \ref{tab-coefficients} in the Appendix.

\subsubsection{Numerical discretization}
The conservation equations are discretized based on a low-dissipative Riemann solver\cite{zhang2017weakly} to increase stability.
The continuity and momentum equations are approximated as
\begin{equation}
	\frac{\text{d} \rho}{\text{d} t} = 2 \rho_i \sum_{j}^{N}  (\mathbf{v_i}-\mathbf{v^*}) \nabla W_{ij} V_j,
	\label{discretize-continuity-equ-riemann}
\end{equation}
\begin{equation}
	\frac{\text{d} \mathbf v}{\text{d} t} = -2\sum_{j}^{N} m_j \frac{P^*}{\rho_i\rho_j} \nabla W_{ij}
	+\frac{2}{\rho_i}\sum_{j}^{N} \widetilde{\mu}_{ij} \frac{\mathbf{v_{ij}}}{r_{ij}}V_j\frac{\partial W_{ij}}{\partial r_{ij}}
	+\mathbf{g}.
	\label{discretize-momentum-p-intermediate}
\end{equation}

Here, $\mathbf{v^*}= U^* \mathbf{e}_{ij}+(\overline{\mathbf{v}}_{ij} - \overline{U}_{ij}\mathbf{e}_{ij})$.
$(\bullet)_{ij} = (\bullet)_i - (\bullet)_j$ refers to the particle-pair difference.
$\overline{(\bullet)}_{ij} = [(\bullet)_i + (\bullet)_j]/2$ means the particle-pair inter average and $\widetilde{(\bullet)}_{ij} = 2(\bullet)_i  (\bullet)_j/[(\bullet)_i + (\bullet)_j]$ refers to the pairwise harmonic average.
$\overline{U}_{ij}$ is the projection of the inter-average velocity $\overline{\mathbf{v}}_{ij}$ along the pairwise direction.
The intermediate velocity and pressure are calculated by
\begin{equation}
	\left\{
	\begin{aligned}
		U^* & = \overline{U}_{ij} + \frac{P_{ij}}{2\rho_0 c_0},               \\
		P^* & = \overline{P}_{ij} + \frac{1}{2} \beta_{ij} \rho_0 c_0 U_{ij},
	\end{aligned}
	\right.
	\label{combined-equations}
\end{equation}
where $\beta_{ij} = \min( \eta \max(\mathbf{v}_{ij} \cdot\mathbf{e}_{ij},0), c_0 )$ is the dissipation limiter, and $\eta=3$ is determined by numerical tests and is used throughout this work.
Please note that the numerical dissipation term in the intermediate-pressure formulation (i.e., the second term on the right-hand side of Eq. \eqref{combined-equations}) is activated only in inviscid flow cases, such as the standing wave, and becomes inactive in viscous flows, including channel flow simulations.

The discretization of the $k$ and $\epsilon$ transport equations involves the approximation of the velocity gradient and the diffusion terms.
The velocity gradient is discretized by \cite{morris1996study}
\begin{equation}
	\nabla \mathbf{v}_i=\sum_{j} \mathbf{v}_{ij} \otimes (\mathbf{B}_i \nabla W_{ij}) V_j.
	\label{velo-grad-equ}
\end{equation}

The discretization of the diffusion terms in the $k$ and $\epsilon$ equations is analogous to that of the viscous term in the momentum equation.
Consequently, the discretized formulations of the two transport equations are written as
\begin{equation}
	\frac{\text{d}  k}{\text{d} t}=G_k -\epsilon
	+\frac{2}{\rho_i}\sum_{j}^{N} \widetilde{(D_{k})}_{ij}  \frac{k_{ij}}{r_{ij}}V_j\frac{\partial W_{ij}}{\partial r_{ij}},
	\label{k-discretized-equ}
\end{equation}

\begin{equation}
	\frac{\text{d} \epsilon}{\text{d} t} = C_1 \frac{\epsilon}{k} G_k -C_2 \frac{\epsilon^2}{k}
	+\frac{2}{\rho_i}\sum_{j}^{N} \widetilde{(D_{\epsilon})}_{ij} \frac{\epsilon_{ij}}{r_{ij}}V_j\frac{\partial W_{ij}}{\partial r_{ij}},
	\label{epsilon-discretized-equ}
\end{equation}
where $G_k$ is the discretized generation term of the turbulent kinetic energy, which equals the double contraction of the Reynolds stress tensor and the mean velocity gradient tensor.

\subsection{The boundary conditions and schemes for stability and efficiency}
For both the laminar and turbulent simulations, the wall boundary condition of the effective pressure gradient term is based on the one-side Riemann solver\cite{zhang2017weakly}.
The transport velocity formulation\cite{zhang2017generalized} is used to avoid the tensile instability and bound the zero-order consistency error, and the duel-criteria time stepping scheme\cite{zhang2020dual} is adopted to increase computational efficiency.

As for the laminar simulation exclusively, the non-slip wall boundary condition is imposed on the viscous term.
While for the turbulent simulation, the step-wise wall function method is adopted as the wall boundary conditions.
The details of the Lagrangian particle-based wall function implementation can be found in Ref. \cite{wang2025weakly}, in which rigorous convergence tests have been done on both velocity and turbulent kinetic energy for the wall-bounded turbulence flows.

Importantly, all the simulations mentioned in this work are based on the Wendland kernel with the smoothing length $h=1.3dp$ by default, where $dp$ is the particle size.
\section{The zero-order consistency residue and background pressure}
\label{sec-residue}
As shown in Eq. \ref{discretize-momentum-p-intermediate}, the pressure gradient term is discretized based on the conservative formulation which adopts the particle-pair inter average instead of the pair-wise difference.
The general conservative particle approximation for an arbitrary scalar $\phi$ is\cite{wu2024sph}
\begin{equation}
	\nabla \phi_i=-\sum_{j} (\phi_i+\phi_j) \nabla W_{ij} V_j .
	\label{eq-conservative-particle-grad}
\end{equation}

The root cause of the zero-order consistency issue lies in Eq. \eqref{eq-conservative-particle-grad}.
This is because when reproducing a constant field, the inter-average part $\phi_i+\phi_j$ can not be zero, furthermore, the kernel-gradient summation will not vanish, as well, due to the irregular particle distribution during simulation.

As a consequence of violating the zero-order consistency, using the conservative gradient operator Eq. \eqref{eq-conservative-particle-grad} to reproduce a constant pressure field may introduce an additional residual term, $\text{P} \sum \nabla W$\cite{kum1995viscous}.
Colagrossi et al. \cite{colagrossi2012particle} later provided a rigorous mathematical derivation, demonstrating the general existence of this term, which leads to the following expression after omitting the first-order truncation error:
\begin{equation}
	(\nabla p)_i
	= \sum_j W_{ij}V_j (\nabla p)_i + 2 p_i \sum_{j} \nabla W_{ij} V_j.
	\label{eq-Colagrissi-residue}
\end{equation}
where $p$ is the pressure, $W$ is the kernel function, and $V$ refers to the volume.

The second term on the right-hand side (RHS) is defined as the zero-order consistency residue in this work.
Besides, the kernel-gradient summation is referred to as the zero-order consistency error, as it should ideally vanish to ensure consistency.
The residual term dissipates the flow because the direction of kernel gradient summation is generally opposite to the flow direction\cite{colagrossi2012particle}.

Since the residual term consists of two parts, (1) the pressure of the particle $i$ and (2) the zero-order consistency error, to reduce the residual term, the first idea is to decrease $p_i$.
Although the particle pressure is determined by the flow field and cannot be directly modified, it can still be reduced by lowering the background pressure, as the pressure considered here is relative.
However, a reduction in background pressure may promote the emergence of negative pressure regions, and hence leads to the tensile instability, deteriorating the stability.
Therefore, this idea is theoretically not feasible.

The second way for decreasing the residue term is to reduce the zero-order consistency error.
Since the kernel gradient summation (consistency error) is directly related to the particle distribution, employing a reliable particle redistribution technique may help control or bound this error.
The early redistributing schemes tend to impose an extra background pressure in the equation of state (EOS)\cite{morris1996study} or the momentum equation\cite{marrone2013accurate} to improve the particle regularity.
Although they contribute to avoiding the negative pressure and keeping the fluid particles uniformly distributed, the increment of background pressure magnifies the residue, causing extra numerical dissipation\cite{adami2013transport}.
In extreme cases, when the background pressure is excessively high, the flow can be significantly suppressed or even completely frozen\cite{kum1995viscous}.

To suppress the consistency error while not increase the background pressure, Adami et al. \cite{adami2013transport} proposed the transport velocity formulation (TVF), where the background pressure is utilized to drive particle relaxation rather than being directly imposed in the momentum equation or the EOS.
Similar to the TVF, the particle shifting technique (PST) \cite{sun2017deltaplus} also avoids introducing additional background pressure.
While different from the TVF that does not modify the particle velocity, the PST updates the particle velocity and may loss the exact momentum conservation\cite{khayyer2017enhancement}.
The latest version of the TVF technique \cite{zhu2021consistency}
is directly correlated with the zero-order consistency error instead of the background pressure.
With the state-of-the-art TVF scheme, the consistency error becomes bounded without introducing additional background pressure, and, theoretically, the zero-order consistency issue should be mitigated.
In practice, for the cases where background pressure is close to zero, such as the flow around a cylinder \cite{wang2025efficient,zhang2023lagrangian,adami2013transport}, this issue does not produce any noticeable adverse effects.

Nevertheless, the background-pressure-related $p_i$ in the residue remains a potential risk for the conservative SPH method, and as discussed in Section \ref{sec-intro}, the zero-order consistency issue may still lead to problems in the two specific scenarios, namely, the free-surface and channel flows.
In the following sections, we will investigate why this issue becomes unmanageable in the two representative flow scenarios, from both theoretical and numerical perspectives, and examine several potential remedies.

\section{Residue behavior in the free-surface flow}
\label{sec-freesurface}
First, we categorize the free-surface flows into two types: with gravity or without gravity.
This is because the excessive numerical dissipation is mostly observed in the cases with gravity, including the standing wave\cite{lyu2023derivation}, wave propagation\cite{zhang2025towards} and oscillating water chamber\cite{wen2018improved} cases.
Moreover, for these cases, merely increasing spatial resolution has almost no effect in reducing the dissipation.
On the contrary, for the cases without gravity, such as the rotating patch, the dissipation-induced energy decay is minor, and merely increasing the resolution significantly reduces the dissipation\cite{ren2023efficient}.

Therefore, we hypothesize that gravity is a key factor contributing to the unexpected numerical dissipation.
The next section begins by examining why excessive dissipation becomes pronounced in the gravity-driven free-surface flow.
Subsequently, using the inviscid standing wave case, we perform a comprehensive investigation into the effects of two key factors, initial wave amplitude and water depth, on the energy dissipation induced by the residue.

\subsection{Theoretical analysis}
The gravity-driven free-surface flow discussed in this work refers to an idealized water flow influenced solely by gravity as the external force, with no consideration of physical dissipation mechanisms that could lead to energy decay.
Consequently, any observed mechanical energy decay can be attributed to numerical dissipation\cite{zhang2025towards,liang2023study}.

As discussed in Eq. \eqref{eq-Colagrissi-residue}, the residual consists of two components: the background pressure term and the consistency error term.
While the latter is strictly bounded, the former remains unconstrained.

\begin{figure}[htb!]
	\centering
	\includegraphics[trim = 12.7cm 0cm 0cm 9.2cm, clip,width=0.8\textwidth]{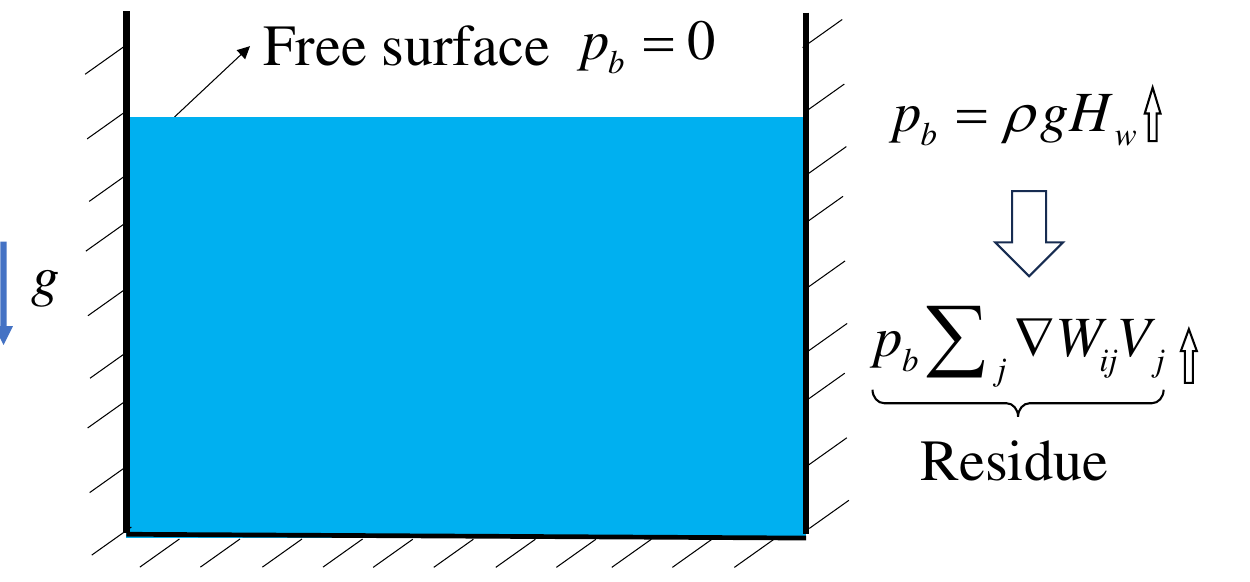}
	\caption
	{
		The flow characteristic of the gravity-driven free-surface flow,
		and with the increase of the water depth $H_w$, the background pressure $p_b$ increases, leading to the increment of the residue.
	}
	\label{fig-static-water-p}
\end{figure}

For free-surface flows driven by gravity, the (background) pressure at the free surface is typically prescribed as zero, serving as the reference pressure throughout the simulation, as shown in Fig. \ref{fig-static-water-p}.
Therefore, all the fluid beneath the free surface is under positive pressure, and according to the hydrostatic pressure equation, $p_b=\rho g H_w$, the background pressure increases with the water depth, $H_w$.
That means, near the bottom of the water tank, the inevitable high background pressure exists, and as denoted in Section \ref{sec-residue}, the high background pressure leads to a large value of the residue, as illustrated in Fig. \ref{fig-static-water-p}.

Theoretically, a deeper water tank is expected to exhibit stronger numerical dissipation due to the relatively higher background pressure near the bottom.
This observation was previously noted by Colagrossi \cite{colagrossi2013smoothed}, although it was not examined in detail.
To numerically test this point, we later test the water depth based on the standing wave case.

Additionally, it is worth noting that the large residue does not directly lead to the high non-physical dissipation, since the residue effects from the momentum equation.
Consequently, the energy decay due to the residue occurs only when the fluid is in motion and ceases once the flow reaches a steady state.
This point will also be tested in the numerical tests when studying the influence of the initial wave amplitude.

%

\subsection{Sensitivity analysis}
In the previous analysis, we identified the background pressure as the primary source of the residue.
This background pressure can be decomposed into two components: a dynamic part (dynamic pressure) and a static part (static pressure).
While both the dynamic and static pressure contribute to the residue, they exhibit distinct characteristics in terms of how they influence the residue-induced energy dissipation.
To investigate these characteristics, numerical tests are conducted in this section to analyze the effects of each component.
The dynamic pressure is related with the initial wave amplitude, and the static pressure is presented by the water depth.

To make reasonable comparisons between different factors, and correctly analyze the sensitivity, the dimensionless is of critical importance.
The simulation time and total mechanical energy are non-dimensionalized as
\begin{equation}
	\begin{cases}
		t^* = t\sqrt{|\mathbf{g}| / L_c}, \\[10pt]
		E^*=\dfrac{E_t-E_0}{E_0-E_{inf}},
	\end{cases}
	\label{eq-dimensionless-free-surface-flow}
\end{equation}
where $L_c$ is the characteristic length, $E$ refers to the total mechanical energy and subscript  $_0$, $_t$, $_{inf}$ denote the values at the initial time, a given time instant $t$, and the infinite time limit, respectively.
The superscript $^*$ refers to the dimensionless values.
Please note that all the numerical tests strictly follow the principle of controlled variables, meaning that in each test, only the variable under investigation is altered while all other parameters are held the same.

\subsubsection{Initial wave amplitude}
The initial wave amplitude is defined in the shape function of the initial free surface, as
\begin{equation}
	\eta_0 = A \cos(\pi x),
	\label{eq:standing-wave-free-surface-shape}
\end{equation}
where $A$ is the initial wave amplitude, $\eta_0$ refers to the initial free surface shape.
As the initial wave amplitude increases, the elevated fluid near the wall generates a stronger dynamic pressure input.

\begin{figure}[htb!]
	\centering
	\includegraphics[trim = 10cm 0cm 0cm 2cm, clip,width=0.75\textwidth]{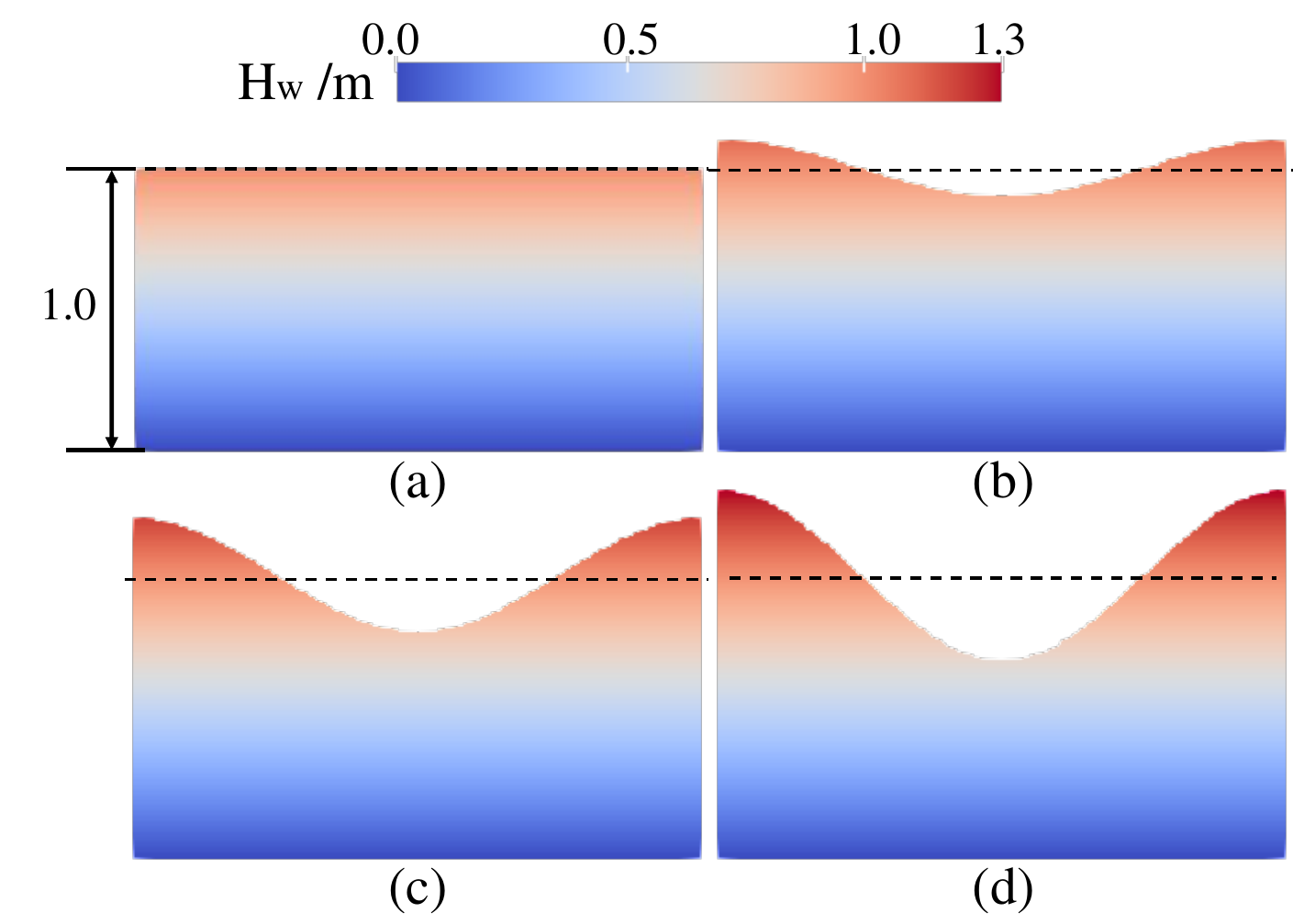}
	\caption
	{
		Standing wave: the initial fluid particle distributions with 4 different initial wave amplitudes,
		(a) 0.0
		(b) 0.1
		(c) 0.2
		(d) 0.3.
		And the black dot lines denote the average water depth.
	}
	\label{fig-wave-amplitude-initial}
\end{figure}

We conducted the tests with four different initial wave amplitudes, as demonstrated in Fig. \ref{fig-wave-amplitude-initial}.
The characteristic length for computing the dimensionless time is the initial wave amplitude.
Besides, for this test only, since modifying the initial wave amplitude, $A$, will not change the $E_{inf}$, we set $E_{inf}=0$ to prevent the possible infinite value when $A=0$.

\begin{figure}[htb!]
	\centering
	\includegraphics[trim = 15.7cm 0cm 0cm 3.5cm, clip,width=0.8\textwidth]{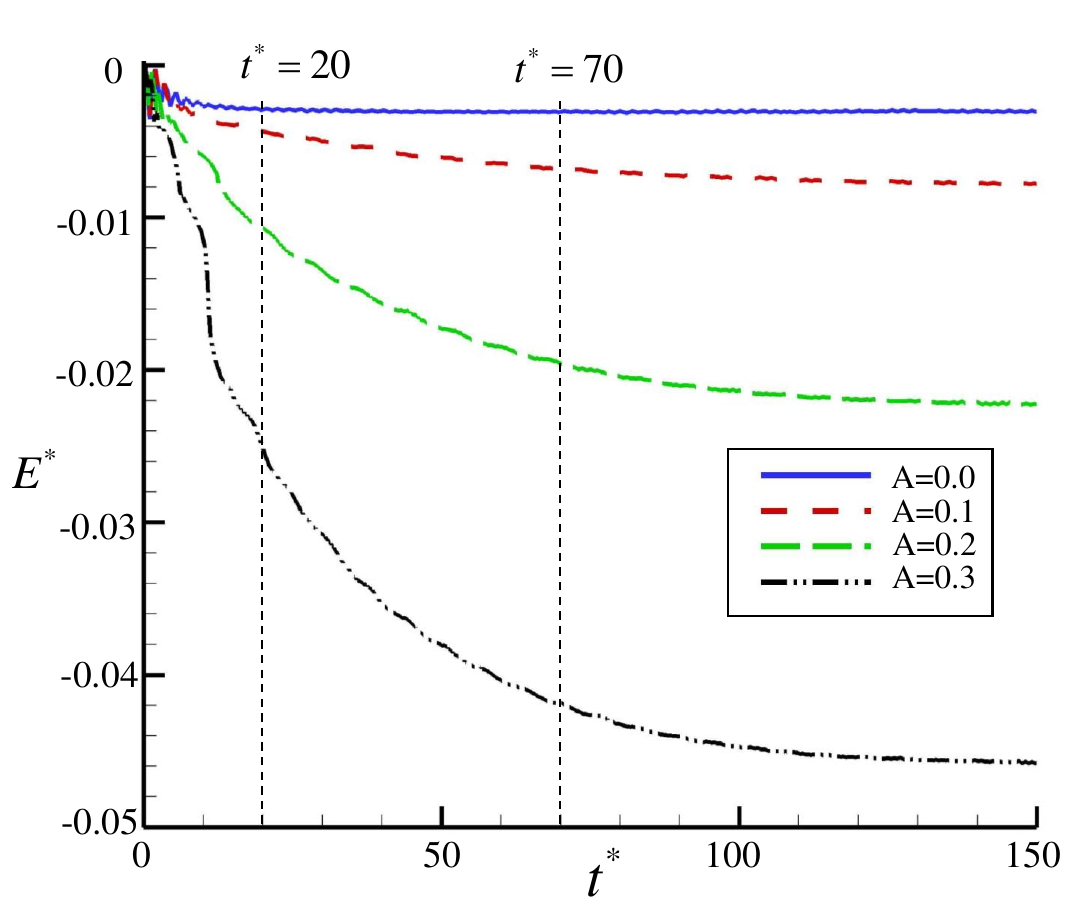}
	\caption
	{
		Standing wave: time evolution of dimensionless total mechanical energy for varying initial wave amplitudes (linear time axis),
		the region enclosed by the dashed lines, $t^* \in [20,70]$, is used for fitting.
	}
	\label{fig-wave-amplitude-result}
\end{figure}

Figure \ref{fig-wave-amplitude-result} shows that the initial wave amplitude has a significant influence on the energy decay.
With the amplitude increasing, the energy decay increases, meaning that the dynamic pressure contributes to the residue-induced numerical dissipation.
Moreover, almost no energy loss is observed when the amplitude is zero (no input dynamic pressure).

\begin{figure}[htb!]
	\centering
	\includegraphics[trim = 15.7cm 0cm 0cm 4.29cm, clip,width=0.8\textwidth]{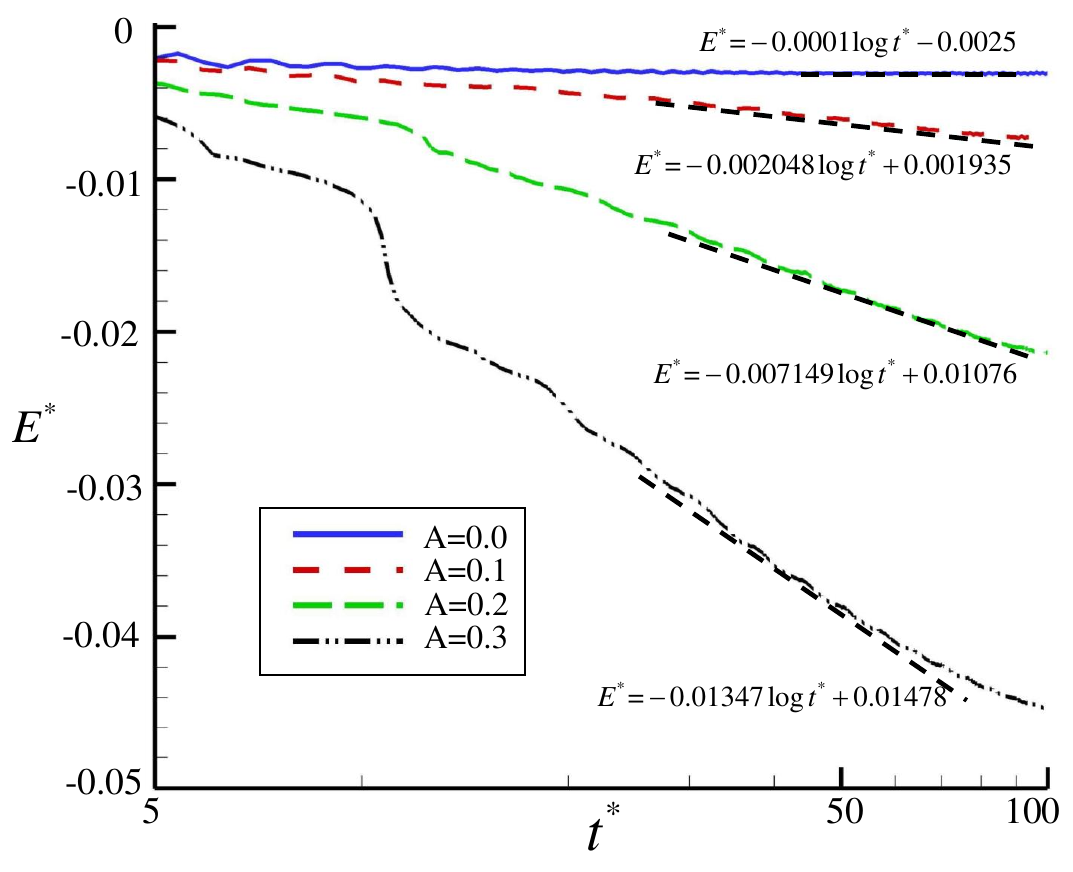}
	\caption
	{
		Standing wave: time evolution of dimensionless total mechanical energy for varying initial wave amplitudes (logarithmic time axis) and the linear fitting results, where the data is obtained for $t^* \in [20,70]$.
	}
	\label{fig-wave-amplitude-result-log}
\end{figure}

To study the energy dissipation rate, we also present the data on a logarithmic timescale, and the time interval, which exhibits the stable energy decay, is selected for linear curve fitting, as shown in Fig \ref{fig-wave-amplitude-result-log}.
The detailed fitting results are presented in table \ref{table-initial-wave-amp}, and it turns out that the energy decay over time caused by the numerical dissipation exhibits a clear exponential law.
As the initial wave amplitude increases, the energy dissipation rate also increases.

\begin{table}
	\scriptsize
	\centering
	\caption{Standing wave: the linear fitting data for different initial wave amplitudes,
		with $R^2$ representing the coefficient of determination to assess fitting quality,
		where values closer to 1 indicate better fitting.}
	\renewcommand{\arraystretch}{1.5}
	\begin{tabular}{l c c c c}
		\hline
		$A$   & 0.0      & 0.1      & 0.2      & 0.3      \\
		\hline
		Slope & -0.00013 & -0.00205 & -0.00715 & -0.01347 \\
		$R^2$ & 0.6991   & 0.9916   & 0.9985   & 0.9973   \\
		\hline
	\end{tabular}
	\label{table-initial-wave-amp}
\end{table}
%
\subsubsection{Water depth}
\begin{figure}[htb!]
	\centering
	\includegraphics[trim = 5.1cm 0cm 0cm 8.9cm, clip,width=1.0\textwidth]{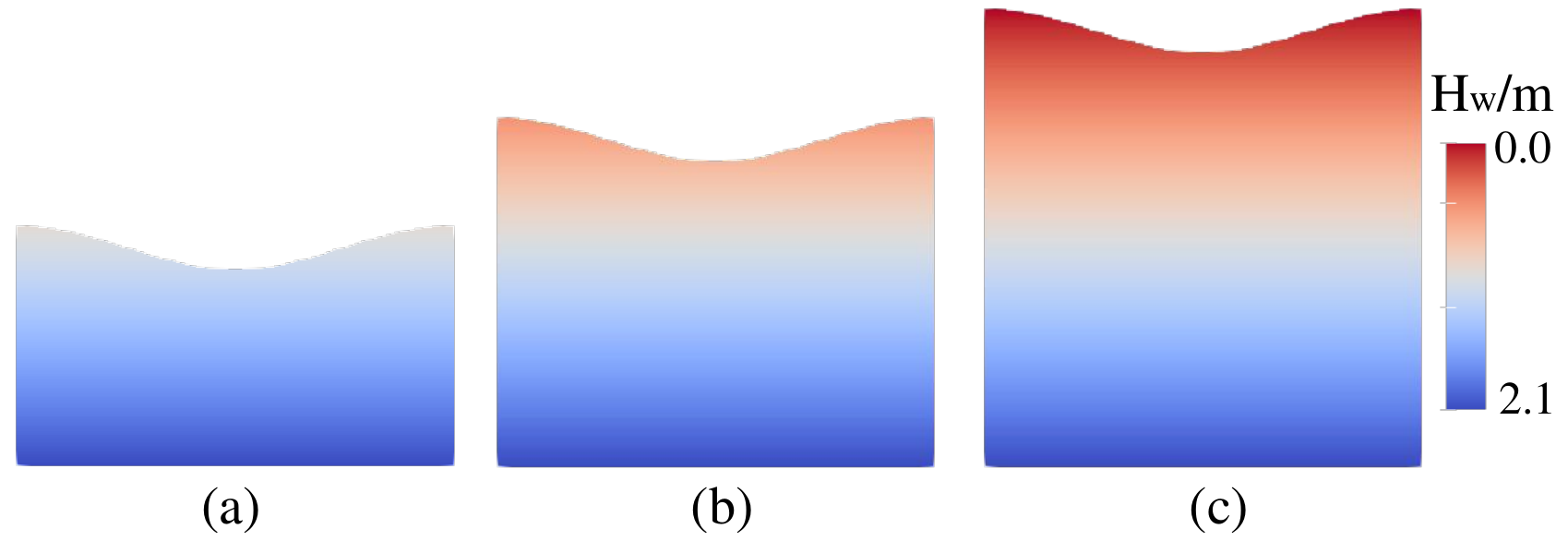}
	\caption
	{
		Standing wave: the initial fluid particle distributions for testing the water depths, $H_w=$
		(a) 1.0
		(b) 1.5
		(c) 2.0.
	}
	\label{fig-wave-depth-initial}
\end{figure}

In the gravity-driven free-surface flow, since the background pressure is directly governed by the water depth, we test the five different depths, ranging from 1.0 to 3.0 with an interval of 0.5.
As illustrated in Fig. \ref{fig-wave-depth-initial},
the water depth here refers to the average height of the water column, and the initial wave amplitude is fixed at 0.1.

The dimensionless time and total mechanical energy are calculated according to Eq. \eqref{eq-dimensionless-free-surface-flow}, in which $E_{inf}=0.5\rho g L_{w} H_{w}$, $L_c=H_{w}$ and $L_{w}$ refers to the width of the water tank.

\begin{figure}[htb!]
	\centering
	\includegraphics[trim = 16.7cm 0cm 0cm 3.45cm, clip,width=0.8\textwidth]{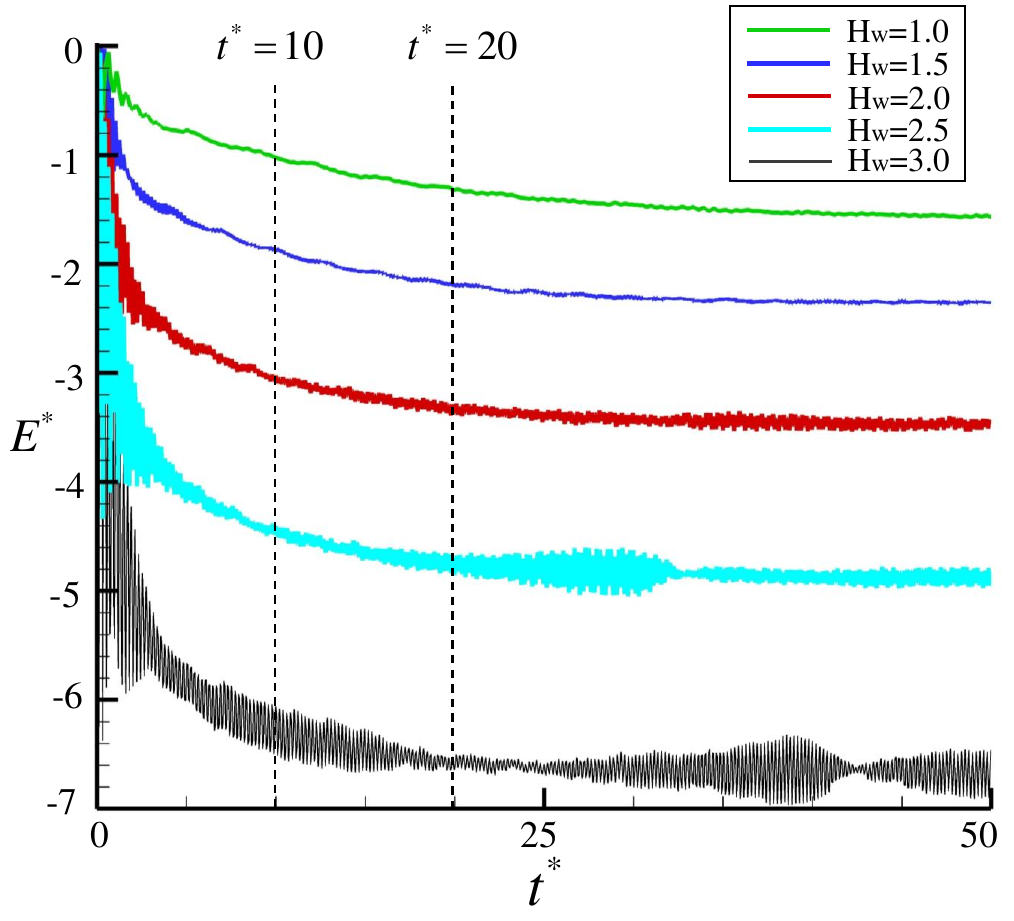}
	\caption
	{
		Standing wave: time evolution of dimensionless total mechanical energy for varying water depths (linear time axis),
		the region enclosed by the dashed lines, $t^* \in [10,20]$, is used for fitting.
	}
	\label{fig-wave-depth-result1}
\end{figure}

\begin{figure}[htb!]
	\centering
	\includegraphics[trim = 16.3cm 0cm 0cm 3.4cm, clip,width=0.8\textwidth]{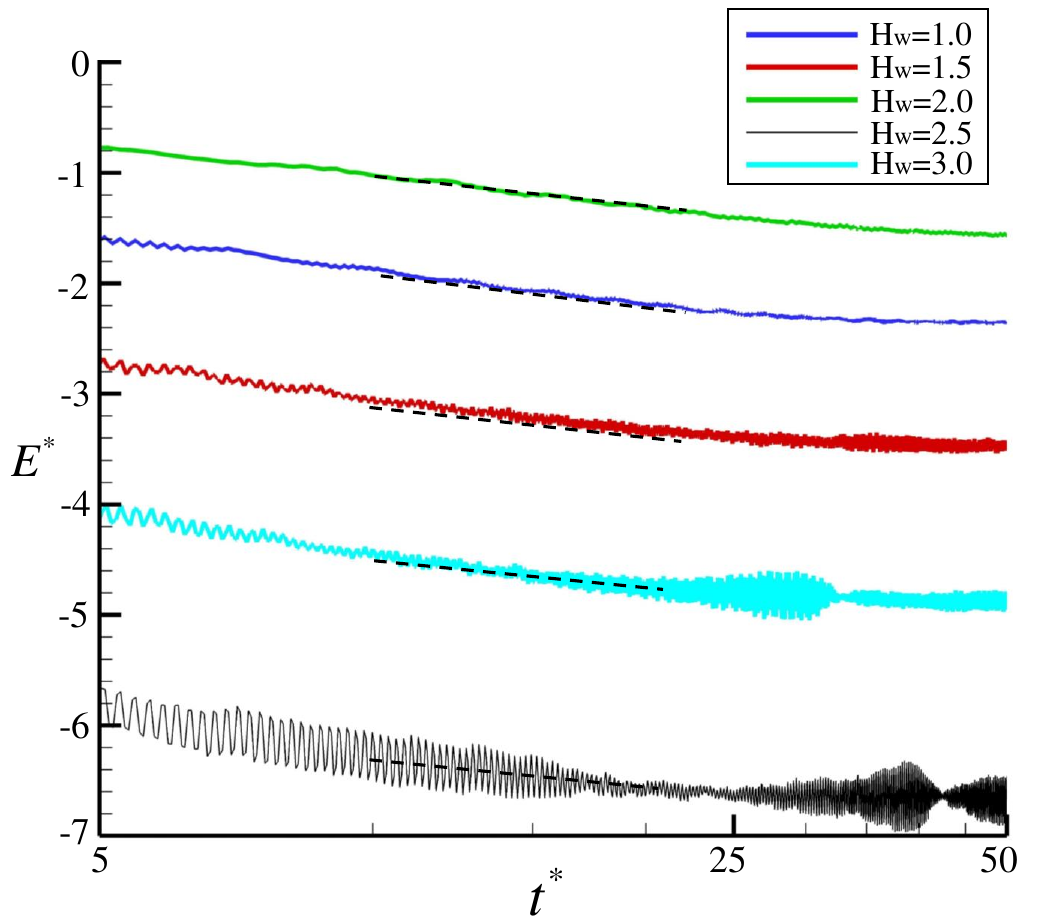}
	\caption
	{
		Standing wave: time evolution of dimensionless total mechanical energy for varying water depths (logarithmic time axis) and the linear fitting results, where the data is obtained for $t^* \in [10,20]$.
	}
	\label{fig-wave-depth-result2}
\end{figure}

Figure \ref{fig-wave-depth-result1} shows that a larger water depth (corresponding to higher background pressure) leads to increased energy dissipation, which agrees with the previous analysis.
To compare the energy dissipation rate, the time axis is plotted on a logarithmic scale, and a time interval, $t^*\in [10,20]$, in which the energy decay is relatively stable, is selected for linear fitting, as presented in Fig. \ref{fig-wave-depth-result2}.

The linear fitting results are summarized in table \ref{table-wave-depth-fitting-data}.
The $R^2$ and MSE illustrate that the residue-induced energy decay of the five cases are well consistent with the logarithmic law and the decrease of the fitting accuracy when $H_w > 2.5$ may be attributed to the numerical oscillation, while the mean squared errors are still relatively small.
The fitted slopes, corresponding to the energy dissipation rates, are in close agreement.

\begin{table}
	\scriptsize
	\centering
	\caption{Standing wave: the linear fitting data for different water depths, where MSE is the mean-squared error.}
	\renewcommand{\arraystretch}{1.5}
	\begin{tabular}{l c c c c c}
		\hline
		$H_w$ & 1.0     & 1.5     & 2.0     & 2.5     & 3.0     \\
		\hline
		Slope & -0.4281 & -0.4401 & -0.4003 & -0.4365 & -0.4189 \\
		$R^2$ & 0.9840  & 0.9784  & 0.8085  & 0.7215  & 0.2951  \\
		$MSE$ & 0.0001  & 0.0002  & 0.0015  & 0.0029  & 0.0163  \\
		\hline
	\end{tabular}
	\label{table-wave-depth-fitting-data}
\end{table}

From Fig. \ref{fig-wave-depth-result2} and Table \ref{table-wave-depth-fitting-data}, the energy dissipation rates under different water depths exhibit a strong similarity.
To provide a theoretical explanation for the similarity in the non-dimensional energy dissipation rates, we derive the corresponding expression from fundamental principles.
Starting from the momentum equation, the residue-induced acceleration change can be written as
\begin{equation}
	\left(\frac{\text{d} \mathbf v}{\text{d} t}\right)_i^R=-\frac{2}{\rho_i}p_{i}\sum_{j} \nabla W_{ij} V_j .
	\label{eq-depth-acc-residue}
\end{equation}
where the superscript $^R$ refers to the residue.

To perform the non-dimensionalisation, we have
\begin{equation}
	\left[\left(\frac{\text{d} \mathbf v}{\text{d} t}\right)_i^R\right]^*=-\frac{2}{\rho_i g H_{wi}}p_{i}\sum_{j} \nabla W_{ij} V_j h ,
	\label{eq-depth-acc-residue-dimensionless}
\end{equation}
where the superscript $^*$ means the dimensionless value and subscript $_i$ refers to the values of the particle $i$.

Note that the uppercase $H_w$ denotes the water depth used for pressure non-dimensionalization, whereas the lowercase $h$ represents the smoothing length used for non-dimensionalizing the zero-order consistency error.
To obtain the energy dissipation rate, we need to time the non-dimensionalized velocity, as
\begin{equation}
	\left[\left(\frac{\text{d} E}{\text{d} t}\right)_i^R\right]^*=-2\frac{p_{i}}{\rho_i g H_{wi}}\mathbf v^*\sum_{j} \nabla W_{ij} V_j h ,
	\label{eq-depth-dEdt-residue-dimensionless}
\end{equation}
where $\mathbf v^*$ is the non-dimensionalized velocity.

Equation \eqref{eq-depth-dEdt-residue-dimensionless} shows that the dimensionless energy dissipation rate is independent of the water depth, $H_w$, because the contributions of the water depth ($\rho_i g H_{wi}$) and the pressure ($p_i$) offset each other based on the hydrostatic pressure equation, in which pressure can be scaled to $\rho g H_w$.
This clearly explains the observed similarity of the dimensionless energy dissipation rates under different water depths, and the dimensionless dissipation rate is merely relevant with the dimensionless velocity, the zero-order consistency error and the smoothing length.
Since the dimensionless velocity can be scaled to the dynamic pressure, Equation \eqref{eq-depth-dEdt-residue-dimensionless} also explains why changing the initial wave amplitude(input dynamic pressure) leads to a corresponding change in the dimensionless energy dissipation rate.

%

\subsubsection{With RKGC technique}
To restore the accuracy and reduce the residue-induced energy dissipation, the kernel gradient correction(KGC)\cite{oger2007improved,zago2021overcoming,ren2023efficient} techniques have been proved to be effective in the standing wave case.
The state-of-the-art reverse KGC scheme\cite{zhang2025towards} not only reduce the nonphysical dissipation but also strictly maintain the momentum conservation.
However, the water depth and initial wave amplitude are limited to 1 and 0.1, respectively, in the RKGC reference\cite{zhang2025towards} for the standing wave case.
In this section, we demonstrate that when the water depth  or the initial wave amplitude becomes sufficiently large, the reverse kernel gradient correction (RKGC) technique fails to completely suppress the energy dissipation.

\begin{figure}[htb!]
	\centering
	\includegraphics[trim = 17.6cm 0cm 0cm 3.5cm, clip,width=0.8\textwidth]{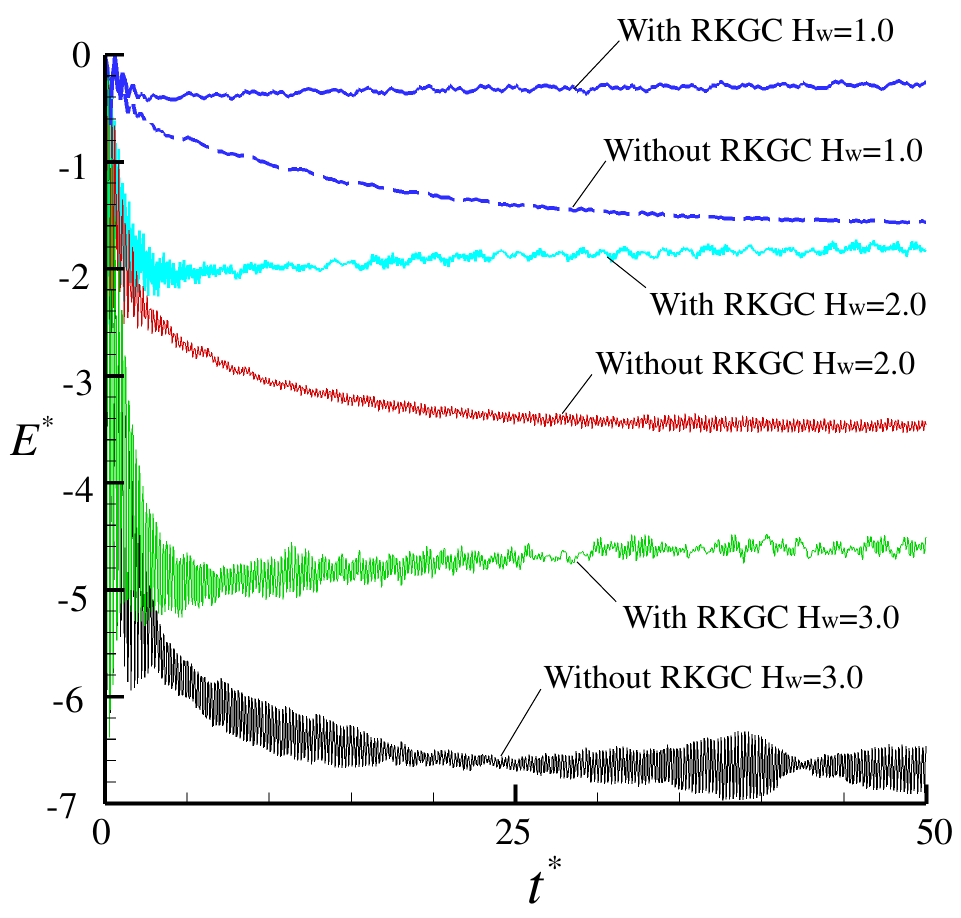}
	\caption
	{
		Standing wave: time evolution of the dimensionless total mechanical energy for varying water depths with or without the RKGC technique.
	}
	\label{fig-standing-wave-RKGC-depth}
\end{figure}

\begin{figure}[htb!]
	\centering
	\includegraphics[trim = 15.77cm 0cm 0cm 4.19cm, clip,width=0.8\textwidth]{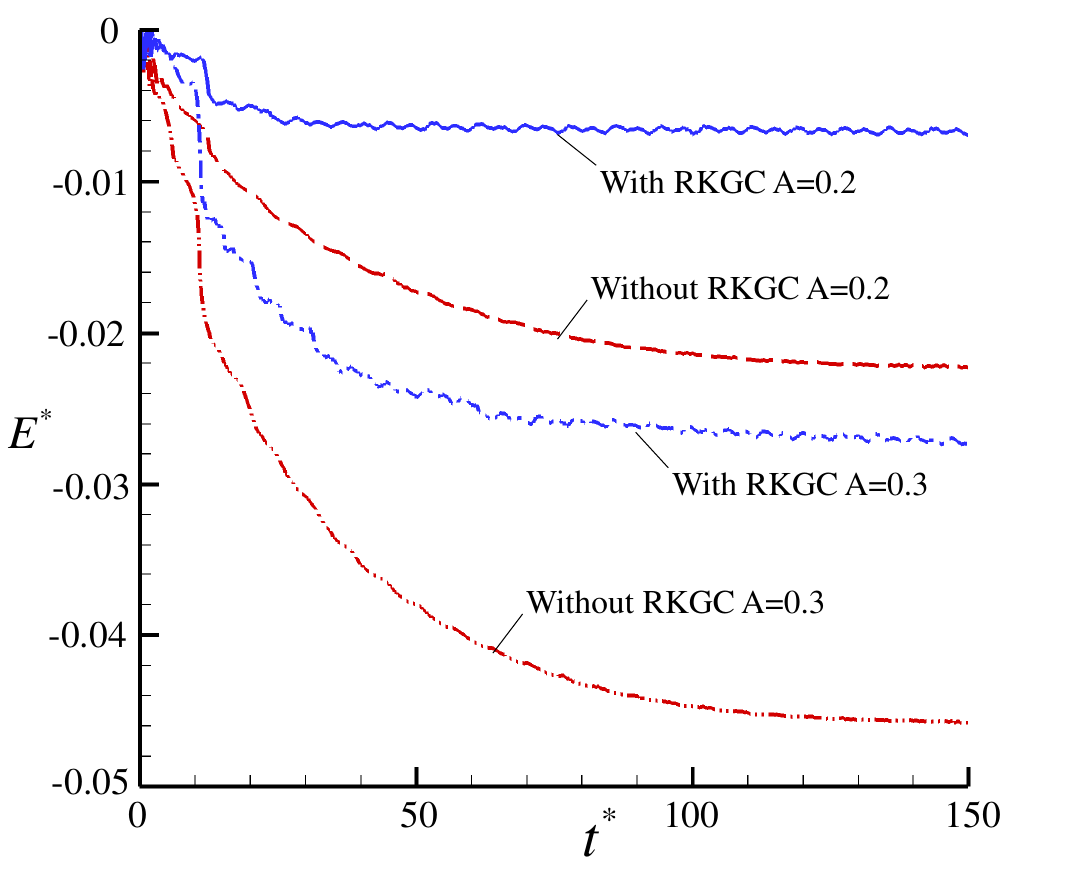}
	\caption
	{
		Standing wave: time evolution of the dimensionless total mechanical energy for varying initial wave amplitudes with or without the RKGC technique.
	}
	\label{fig-standing-wave-RKGC-amp}
\end{figure}

The results with and without the RKGC scheme, under varying water depths ($H_w$) and initial wave amplitudes ($A$), are presented in Fig. \ref{fig-standing-wave-RKGC-depth} and Fig. \ref{fig-standing-wave-RKGC-amp}, respectively.
When the water depth is $H_w=1$ and the initial wave amplitude is $A=0.1$, the RKGC technique effectively suppresses energy decay.
While, as either $H_w$ or $A$ increases, the effectiveness of the scheme decreases.
This is consistent with the previous analysis and demonstrates that the RKGC still has limitations, highlighting the need for caution when employing the conservative SPH method in simulating gravity-driven free-surface flows, especially to avoid conditions involving significant background pressure and strong dynamic pressure.
However, it should be noted that for the cases with very large water depths, some suitable simplifications, such as linear wave or potential flow theories, may be helpful for avoiding the zero-order residue problem.

%
\section{Residual Behavior in the channel flow}
\label{section-channel-error}
As shown in Fig. \ref{fig-intro-damp}, we find that the velocity is damped in a very long channel with the VIPO boundary conditions, especially near the inlet.
We initially suspect that this issue arises from the residue, as the residue can induce additional numerical dissipation.
In this section, we firstly analysis this problem from a theoretical perspective and then conduct numerical tests for validation.

\subsection{Theoretical analysis}
\label{straight-theory-analysis}
\begin{figure}[htb!]
	\centering
	\includegraphics[trim = 8.9cm 0cm 0cm 8.3cm, clip,width=1.0\textwidth]{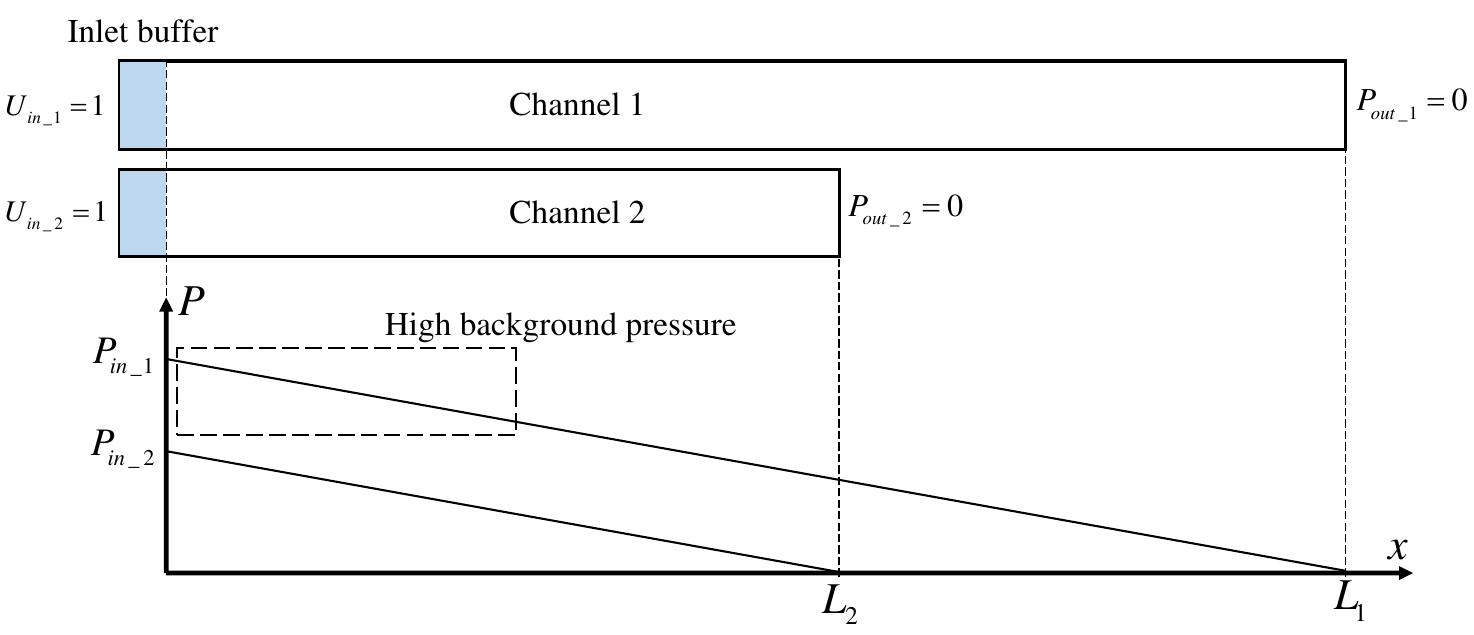}
	\caption
	{
		Theoretical pressure distribution along the main-stream($x$) direction in the two channels with different lengths.
	}
	\label{fig-channel-property}
\end{figure}

Similar to the analysis of the free-surface flow, although the zero-order consistency error is bounded by using the TVF technique\cite{zhu2021consistency}, the background pressure near the inlet is positively correlated with the channel length, as the outlet pressure is usually calibrated to zero in fluid engineering.
Figure \ref{fig-channel-property} further demonstrates this problem, for the two channels with the same VIPO boundary conditions but different channel lengths ($L_1 \gg L_2$), since the pressure gradients are the same, the inlet pressure of the channel $1$ will be much larger than the channel $2$, leading to a high background pressure region near the inlet of the channel $1$.
The higher background pressure leads to the large residue, which results in the numerical dissipation and velocity damping.

To further reveal the underlying mechanism of damping, we analyze the difference between the residue and the pressure gradient terms, as they exhibit an intrinsic opposing relationship in pressure-driven flows.
This comparison is essential because only when the residue dominates over the pressure gradient does the overall flow begin to exhibit noticeable dissipative behavior.
When these two terms reach a balance, we obtain:
\begin{equation}
	(\nabla p)_i -  2 p_i \sum_{j} \nabla W_{ij} V_j =0 .
	\label{eq-residue-compare-with-pressure-gradient}
\end{equation}

In a fully-developed two-dimensional VIPO channel, the pressure gradient is constant everywhere and hence can be expressed as
\begin{equation}
	\nabla p = \frac{P_{in}-P_{out}}{L},
	\label{eq-pressure-gradient-in-channel}
\end{equation}
where $L$ is the channel length and $P_{in}$ stands for the inlet pressure and $P_{out}$ refers to the outlet pressure.
Since the pressure we discuss here is relative pressure, the outlet pressure can be calibrated to zero.
Besides, the pressure in a fully-developed channel can be computed as $P(x) = P_{in}(-x/L+1)$ .
Considering the above two conditions, we find Eq. \eqref{eq-residue-compare-with-pressure-gradient} can be expressed as
\begin{equation}
	\frac{P_{in}}{L} -  2P_{in}(1-\frac{x}{L}) \sum_{j} \nabla W_{ij} V_j =0 .
	\label{eq-residue-compare-with-pressure-gradient-simplified}
\end{equation}

Since $P_{in}$ is a constant, finally, we arrive at
\begin{equation}
	\frac{1}{L} -  2(1-\frac{x}{L}) \sum_{j} \nabla W_{ij} V_j =0 .
	\label{eq-residue-compare-with-pressure-gradient-simplified2}
\end{equation}

Note that Eq. \eqref{eq-residue-compare-with-pressure-gradient-simplified2} holds generally throughout the entire channel, and hence we can assume at the mid-cross section of the channel, we have
\begin{equation}
	\frac{1}{L} -  \sum_{j} \nabla W_{ij} V_j =0 .
	\label{eq-residue-compare-with-pressure-gradient-simplified3}
\end{equation}

Equation \eqref{eq-residue-compare-with-pressure-gradient-simplified3} reveals that whether the residue dominates or not depends exclusively on two factors, the channel length and the zero-order consistency error.
Since the consistency error is bounded by the TVF, the damping effect is positively correlated with the channel length, and the fundamental cause is the excessive background pressure induced by the long channel length.
\subsection{Sensitivity analysis}
\subsubsection{Background pressure}
\label{sec-channel-background}
To validate the above analysis that the excessive numerical dissipation originates from high background pressure, a straightforward treatment is to eliminate this pressure by using the periodic boundary condition instead of the VIPO boundaries.
A constant external body force is used to drive the flow instead of the pressure difference, so that the high background pressure near inlet is avoided.
The simulation settings are concluded in Table \ref{tab-channel-periodic-settings}, the Reynolds number is based on the channel height and mean bulk velocity.
The number of fluid particles across the cross-section is 20.
Please note that the parameters are dimensionless, and the only difference compared with the simulation presented in Fig. \ref{fig-intro-damp} is the boundary condition.
\begin{table}
	\scriptsize
	\centering
	\caption{Laminar channel flow with periodic boundary condition: the simulation settings.}
	\renewcommand{\arraystretch}{1.5}
	\begin{tabular}{lc}
		\toprule
		Parameter                    & Value \\
		\midrule
		Length-to-height ratio $L/H$ & 60    \\
		Body acceleration            & 0.03  \\
		Kinematic viscosity $\nu$    & 10    \\
		Reynolds number $Re$         & 200   \\
		\bottomrule
	\end{tabular}
	\label{tab-channel-periodic-settings}
\end{table}

\begin{figure}[htb!]
	\centering
	\includegraphics[trim = 0cm 0cm 0cm 7.95cm, clip,width=1.0\textwidth]{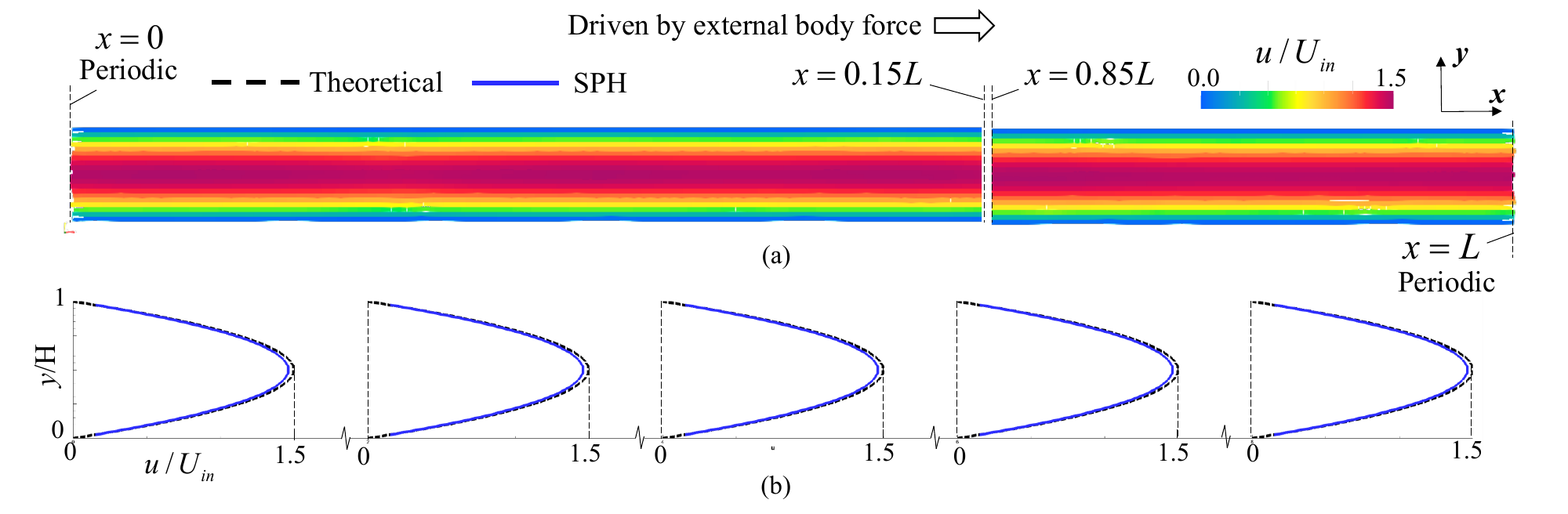}
	\caption
	{
		Laminar channel flow with periodic boundary condition:
		(a) the velocity contour and (b) cross-sectional velocity profiles.
		The five cross-sectional velocity profiles are monitored at positions $x/L=$ 0, 0.25, 0.5, 0.75.
		Some labels are omitted for simplicity.
		The segments where $x\in(0.15L,0.85L)$ are not shown.
	}
	\label{fig-laminar-channel-periodic}
\end{figure}

The results are shown in Fig. \ref{fig-laminar-channel-periodic}.
Compared with the results shown in Fig. \ref{fig-intro-damp}, the unexpected damping clearly disappears, and the cross-sectional velocity profiles agree well with the theoretical values, validating the above analysis.
However, although the use of periodic boundary conditions can effectively bypass the zero-order consistency issue, it is worth noting that such conditions are not commonly used in engineering practice due to the complexity of real-world flow configurations, especially when compared to the widespread adoption of VIPO boundaries.

Please also note that after this section, all the simulations of the channel flow are based on the VIPO boundary conditions.
%
\subsubsection{Channel length}
\label{subsubsection-Channel-length}

\begin{figure}[htb!]
	\centering
	\includegraphics[trim = 0cm 0cm 0cm 0cm, clip,width=1.0\textwidth]{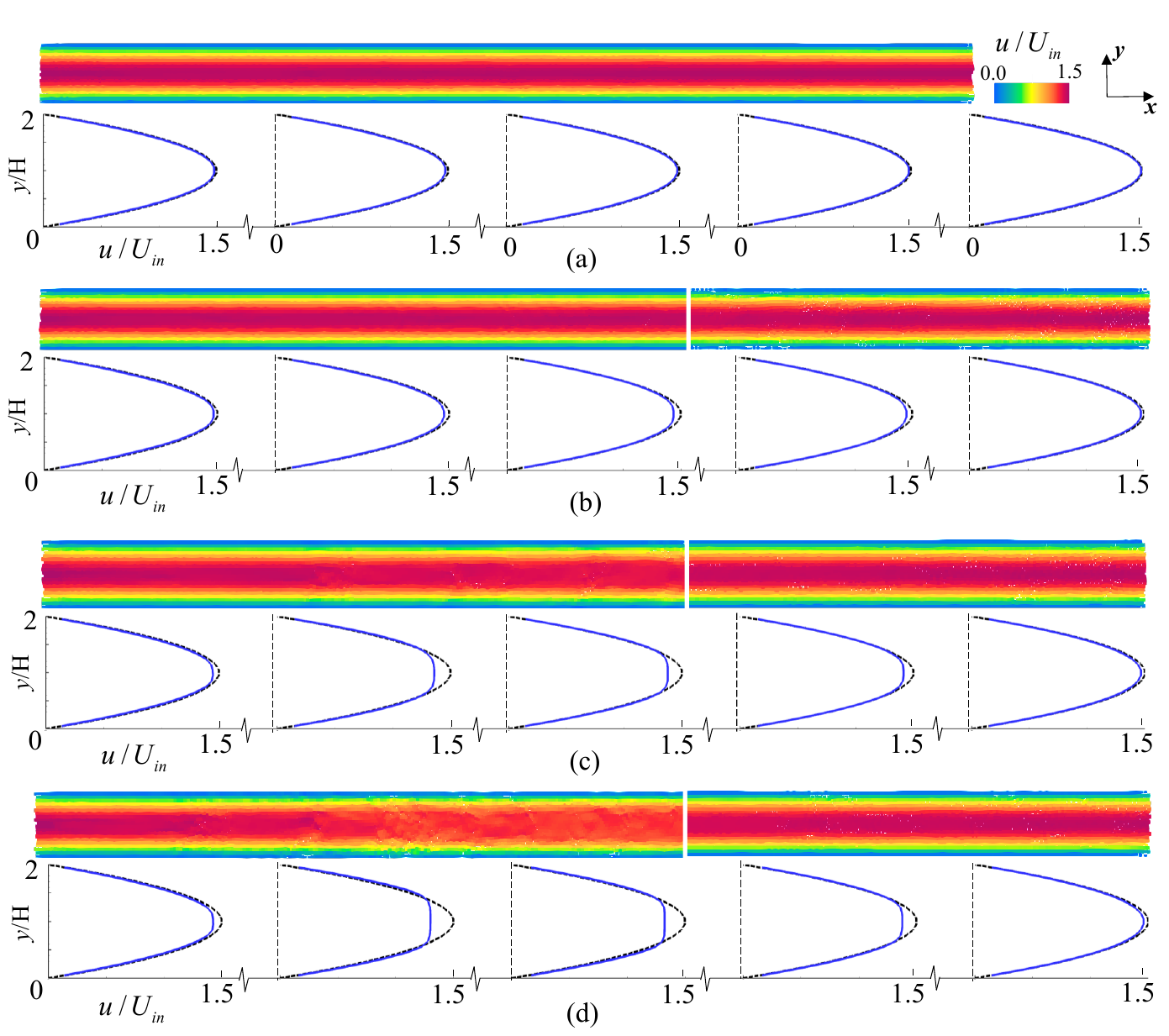}
	\caption
	{
		Laminar channel flow:
		the velocity contours with different length-to-height ratios ($\delta=$ (a) 15, (b) 30, (c) 45 and (d) 60 from up to bottom).
		Each case has five cross-sectional velocity profiles monitored at positions $x/L=$ 0, 0.25, 0.5, 0.75.
		Some labels are omitted for simplicity.
		The dot lines represent the theoretical velocity profile and the blue solid lines are the results calculated by the conservative WCSPH method.
		For the cases where $\delta=$30, 45 and 60, the segments where $x\in(0.15L,0.85L)$ are not shown.
	}
	\label{fig-laminar-vel-contours-different-lengths}
\end{figure}

To study the relation between the damping effect and the channel length, we simulate the laminar straight channel cases with different channel lengths.
The Reynolds number, based on the channel height and mean bulk velocity, is 200, and the fully-developed (parabolic) velocity inlet and zero pressure outlet boundary conditions are used.
The number of fluid particles across the cross-section is 20.
Four different ratios of the channel lengths to the heights, termed as $\delta$, is tested.
The velocity contours together with the cross-sectional velocity profiles calculated by the conservative WCSPH method are shown in Fig. \ref{fig-laminar-vel-contours-different-lengths}.

\begin{figure}[htb!]
	\centering
	\includegraphics[trim = 0cm 0cm 0cm 8.5cm, clip,width=1.0\textwidth]{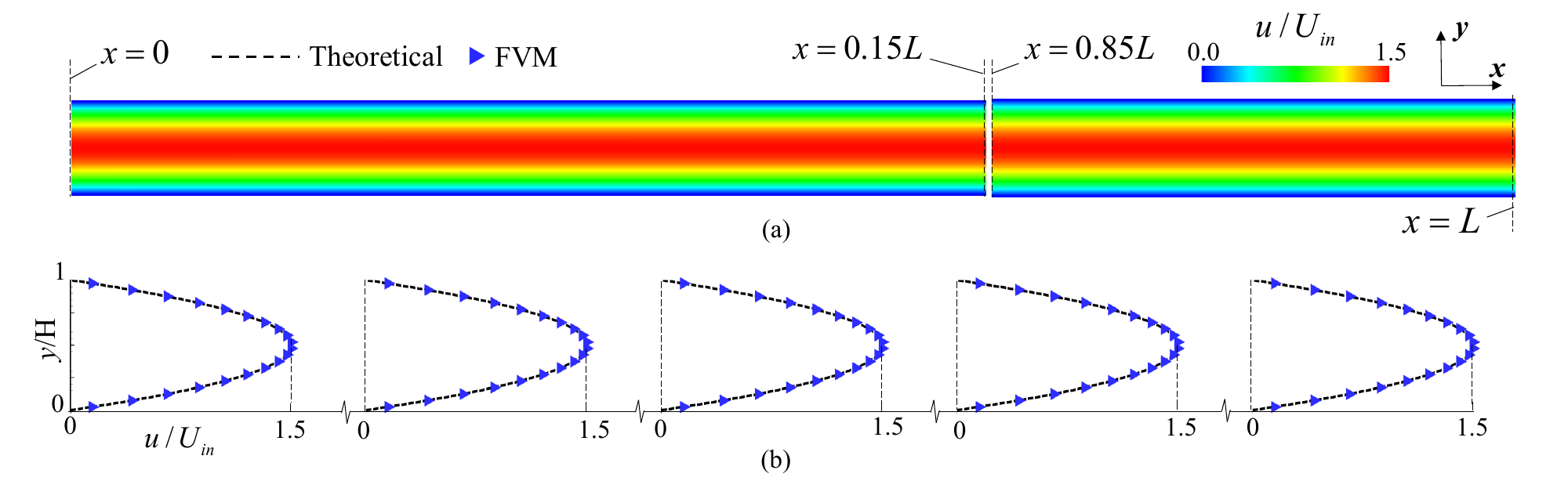}
	\caption
	{
		Laminar channel flow:
		The velocity contour computed by the finite volume method under the same simulation settings as the WCSPH method.
	}
	\label{fig-vel-laminar-fvm}
\end{figure}

Since the analytical (parabolic) velocity profile is imposed at the inlet, after achieving the steady state, the parabolic velocity profile should appear at any cross-section of the channel, as shown in the results calculated by the finite volume method with the same conditions(Fig. \ref{fig-vel-laminar-fvm}).
However, as for the WCSPH results, when $\delta>45$, velocity near inlet is obviously damped due to the residue, and the maximum velocity suffers a loss compared with the theoretical value.
This loss is positively correlated with the channel length, which agrees with the previous analysis, and can be up to 13.4\% when $\delta=60$.

\begin{figure}[htb!]
	\centering
	\includegraphics[trim = 0.6cm 0cm 0.7cm 1.5cm, clip,width=1.0\textwidth]{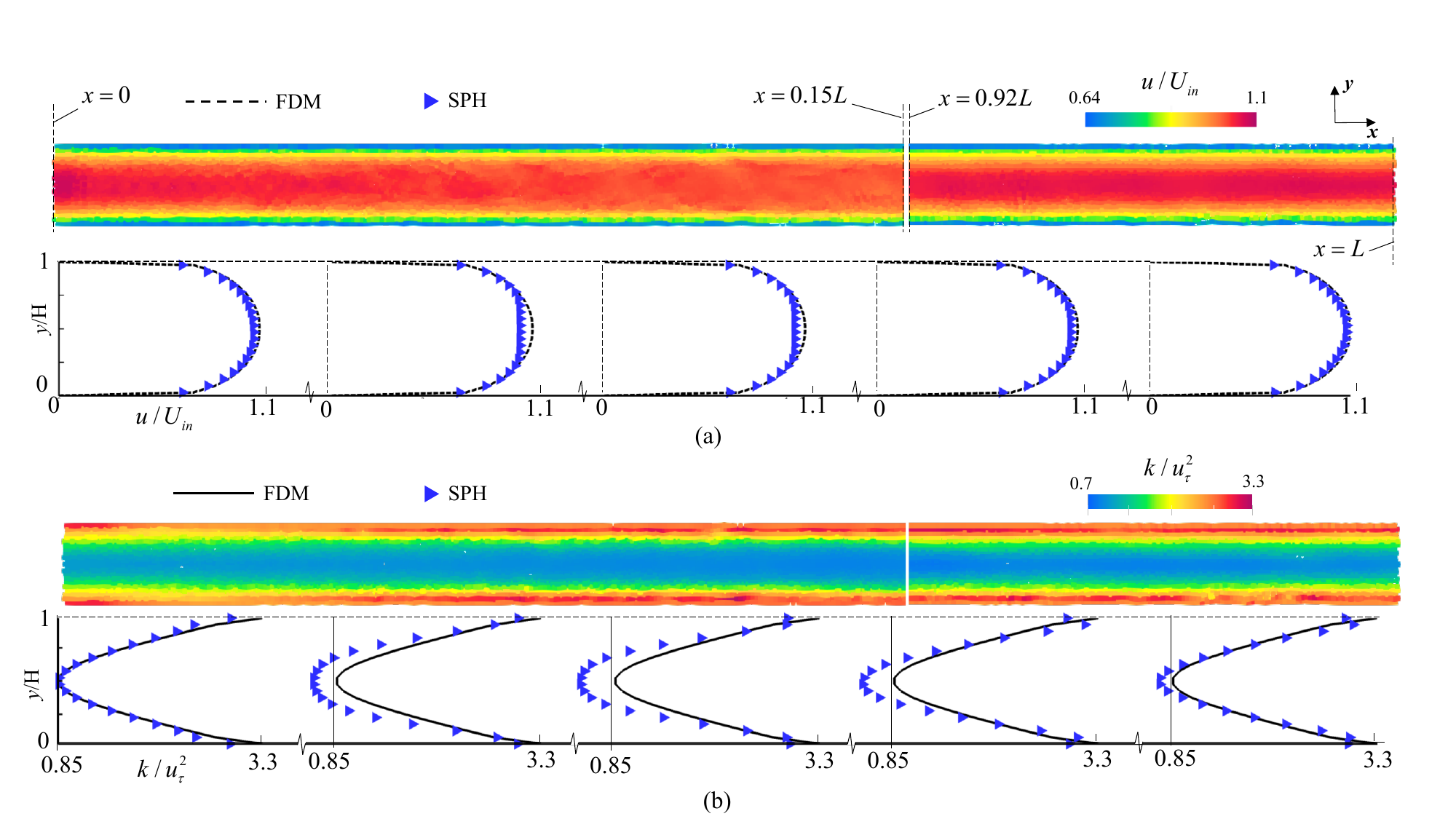}
	\caption
	{
		Turbulent channel flow:
		the contours of the turbulent kinetic energy and velocity together with the cross-sectional data in the long turbulent channel ($\delta=60$, $Re=20000$).
		The cross-sectional profiles are monitored at positions $x/L=$ 0, 0.25, 0.5, 0.75.
		The dot lines represent the FDM results and the blue solid lines are the results calculated by the WCSPH-RANS method without RKGC.
	}
	\label{fig-vel-k-C-WCSPH}
\end{figure}

\begin{figure}[htb!]
	\centering
	\includegraphics[trim = 0cm 0cm 0cm 9cm, clip,width=1.0\textwidth]{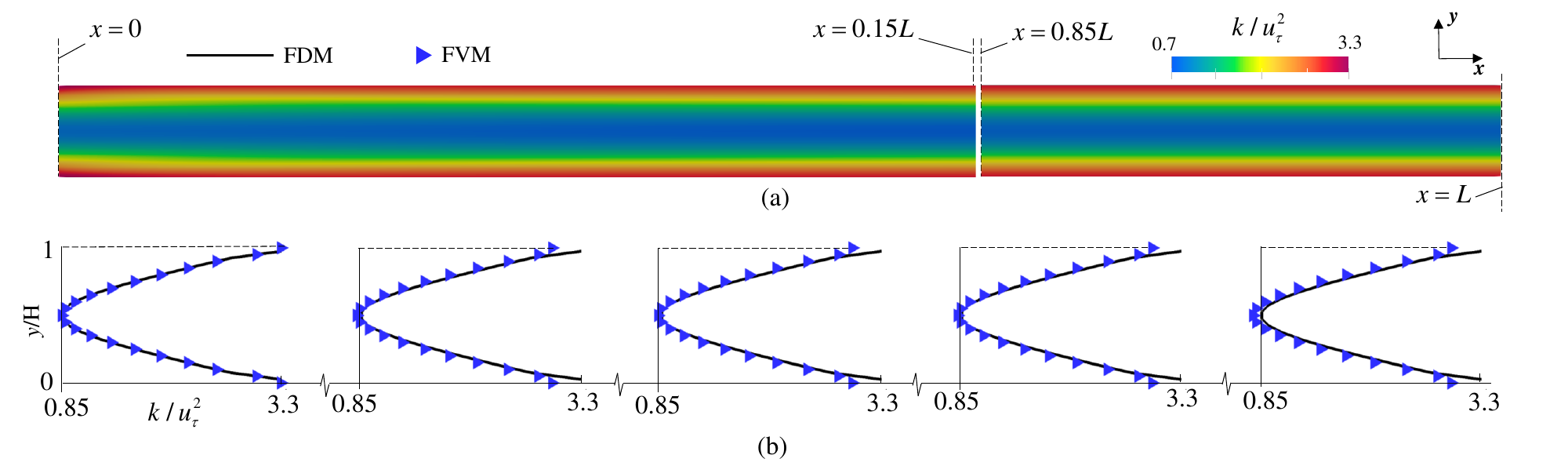}
	\caption
	{
		Turbulent channel flow:
		the velocity contours and the cross-sectional profiles of the laminar channels at positions $x/L=$ 0, 0.25, 0.5, 0.75 calculated by the finite volume method.
	}
	\label{fig-k-fvm}
\end{figure}

Note that this problem also exists in the turbulent simulation with the WCSPH-RANS method\cite{wang2025weakly}.
Figure \ref{fig-vel-k-C-WCSPH} shows the results of a long straight turbulent channel ($\delta=60$) simulated by the WCSPH-RANS method.
The Reynolds number is 20000 and the analytical data obtained\cite{wang2025weakly} from the finite difference method are imposed as the inflow boundary condition, as well as the zero pressure outlet boundary condition.
Compared with the results of the finite difference method (FDM), although the mean flow velocity profiles are not significantly damped, the turbulent kinetic energy profiles near the inlet suffer a clear under-prediction.
We also simulated this long turbulent channel by the finite volume method (FVM) with the exact same conditions, as shown in Fig. \ref{fig-k-fvm}.
The turbulent kinetic energy profiles obtained by the FVM agree well with the FDM results and no obvious under-prediction is observed near the inlet.

%

\subsubsection{Outlet pressure}

As previously discussed, the large residue results from the high background pressure near inlet, therefore, a straightforward idea for reducing the residue effect is to decrease the background pressure.
In this section, we investigate the influence of adjusting background pressure on residue-induced velocity loss.

The background pressure is controlled by changing the outlet pressure.
Note that, for the simulation of the incompressible flow, the pressure here refers to the gauge pressure, and the flow is motivated by the pressure difference.
Therefore, changing the magnitude of the outlet pressure does not affect the physical flow properties, such as the Reynolds number, since the inlet pressure will automatically match the outlet pressure, forming the same pressure difference.

\begin{figure}[htb!]
	\centering
	\includegraphics[trim = 4.4cm 0cm 0cm 6.3cm, clip,width=1.0\textwidth]{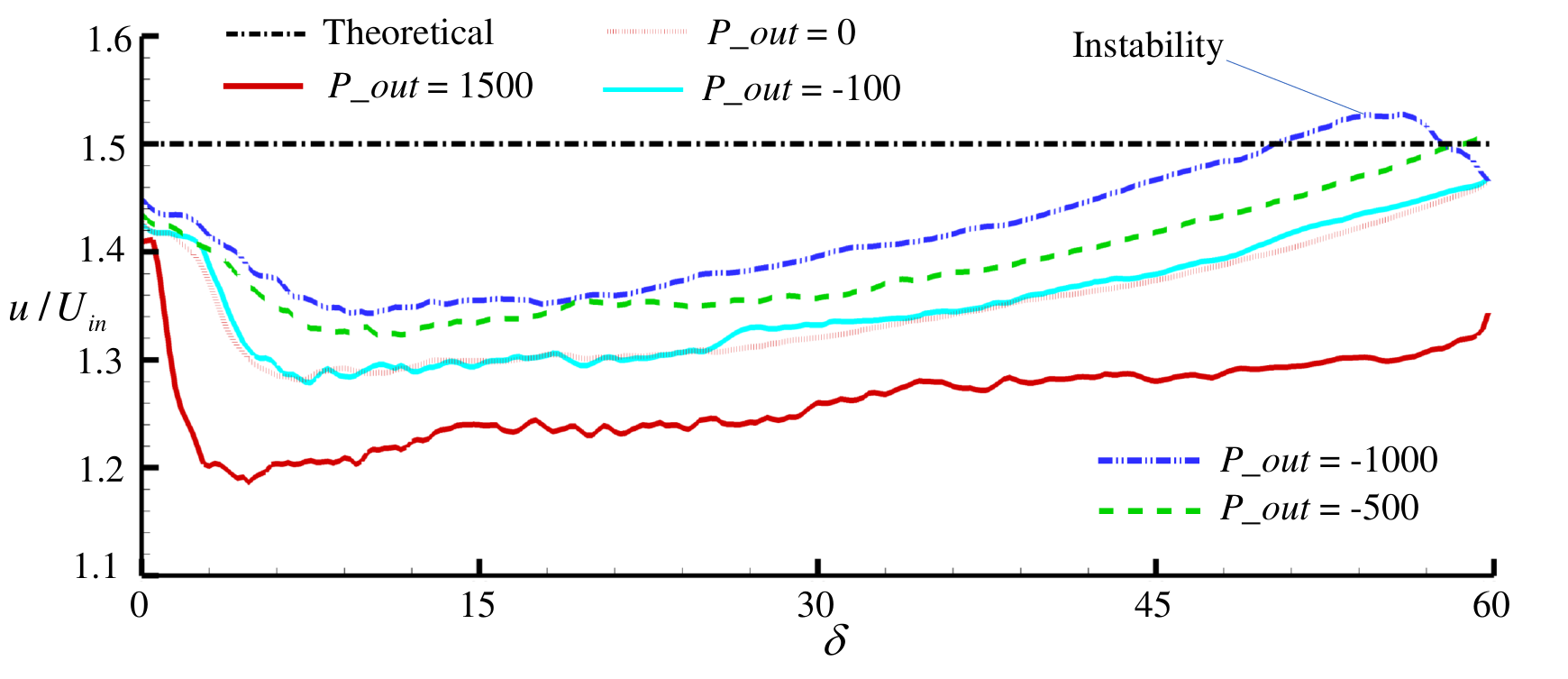}
	\caption
	{
		Laminar channel flow:
		the centerline velocity profiles of the long laminar channel ($\delta=60$) calculated by the WCSPH method at the four outlet pressure ($p_{out}=1500, -100, -500, -1000$).
	}
	\label{fig-adjust-outlet-pressure-combine}
\end{figure}

\begin{figure}[htb!]
	\centering
	\includegraphics[trim = 0cm 0cm 0cm 10.8cm, clip,width=1.0\textwidth]{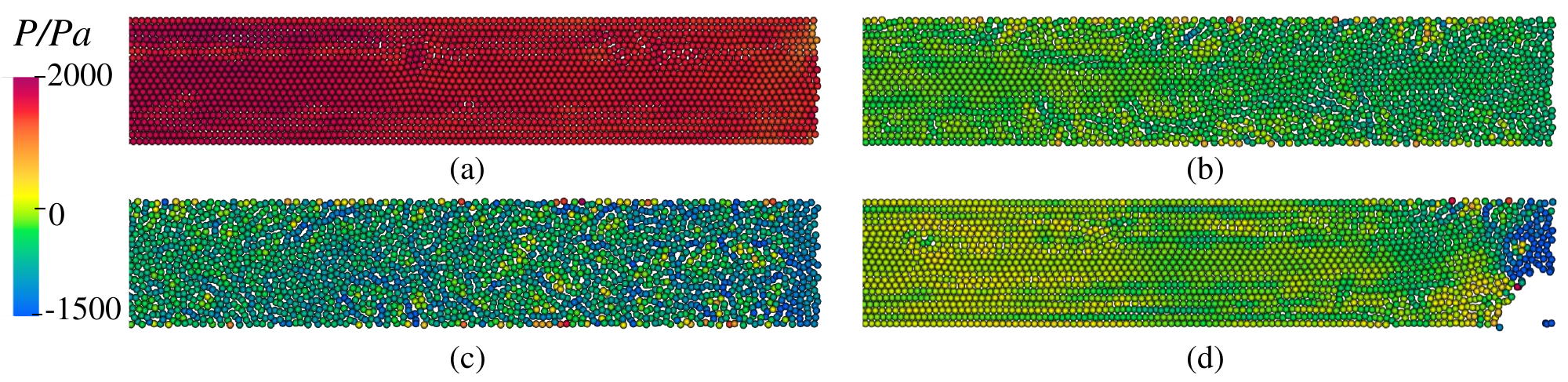}
	\caption
	{
		Laminar channel flow:
		The pressure contours near outlet $[0.92L,L]$  of the long laminar channel ($\delta=60$) calculated by the WCSPH method at the four outlet pressure ($p_{out}=(a) 1500, (b) -500, (c) -1000, (d) -1500$).
	}
	\label{fig-adjust-outlet-pressure-contour}
\end{figure}

The simulation settings are the same as the long ($\delta=60$) straight laminar channel case reported in Sec. \ref{subsubsection-Channel-length}.
Five sets of the outlet pressure, -1000, -500, -100, 0, 3000, are tested.
Figure \ref{fig-adjust-outlet-pressure-combine} shows that with the outlet pressure decreasing, the centerline velocity gradually increases to the theoretical value.
However, when $p_{out}<-1000$, the strong negative outlet pressure exceeds the correction capacity of the transport velocity formulation, and causes the tensile instability, as shown in Fig. \ref{fig-adjust-outlet-pressure-contour}.
Further reduction of the outlet pressure leads to particle splashing near the outlet region, as illustrated in Fig.~\ref{fig-adjust-outlet-pressure-contour}(d).

On the contrary, as shown in Fig. \ref{fig-adjust-outlet-pressure-combine}, imposing a strong pressure at outlet ($p_{out}=1500$) significantly exacerbates the max velocity loss, although the particle distribution becomes quite uniform.
This is also one of the main drawbacks of the background pressure technique\cite{morris1996study,marrone2013accurate}, which is usually adopted to regularization particles at the early age.

The above results demonstrate that, in the conservative WCSPH method, careful selection of the background (or outlet) pressure is essential, as it is closely related to the zero-order consistency residue, which may lead to unintended velocity loss and increased numerical dissipation.

%

\subsubsection{Resolution}

Aiming at reducing the above-mentioned velocity damp in the long laminar channel simulation, we test the influence of the spatial resolution, since it is generally an effective way to reduce errors by increasing the resolution.
The laminar long straight channel ($\delta=60$) is simulated by the conservative WCSPH with three different resolutions, and the simulation parameters are identical to those described in the previous section.
Since the zero-order residue obviously affects the centerline velocity, we observe the profile of the centerline velocity along the flow direction, $x$.

\begin{figure}[htb!]
	\centering
	\includegraphics[trim = 5.7cm 0cm 0cm 6.1cm, clip,width=1.0\textwidth]{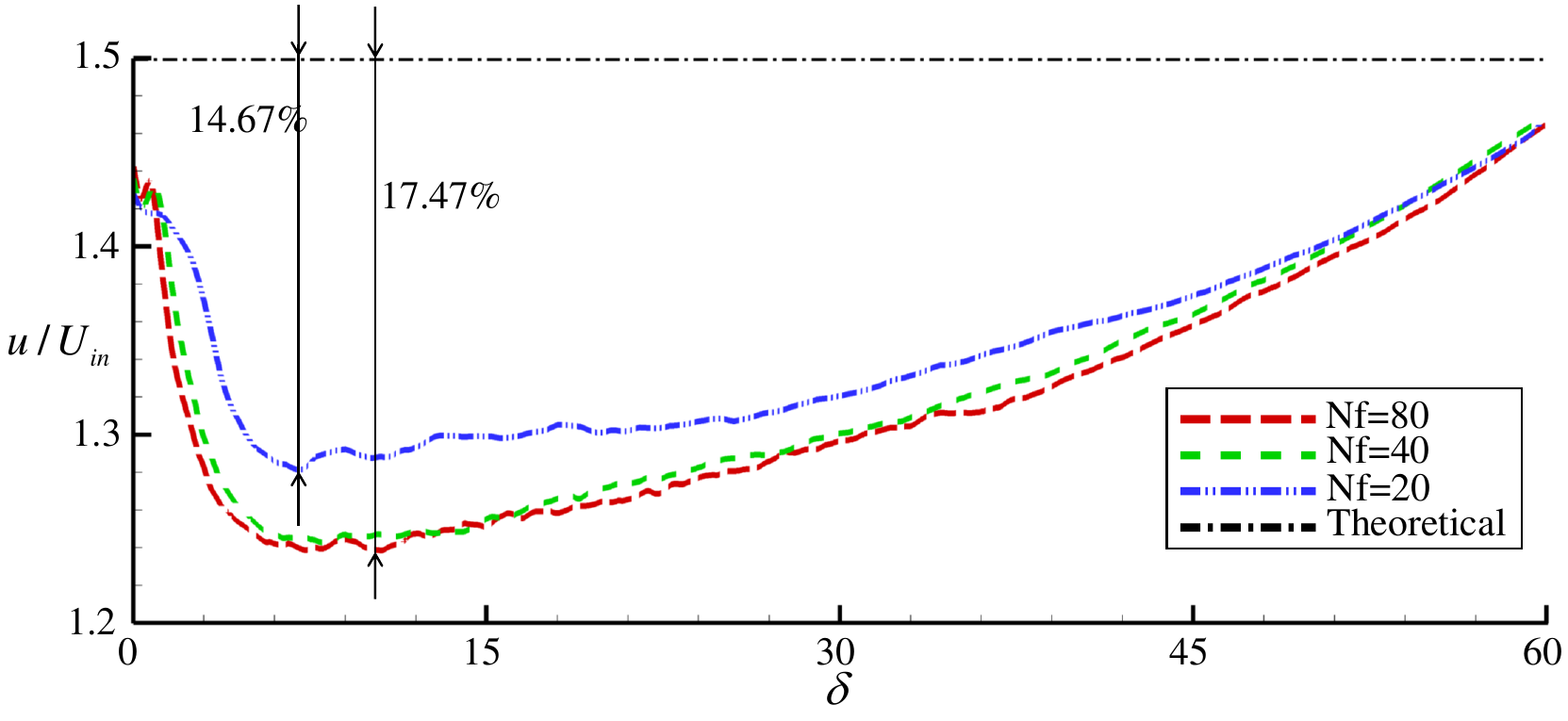}
	\caption
	{
		Laminar channel flow:
		The centerline velocity of the long laminar channels ($\delta=60$) calculated by the conservative WCSPH method at the three resolutions ($N_f=20, 40, 80$, where $N_f$ is the number of fluid particles across the cross-section)
	}
	\label{fig-increase-resolution-centerline-vel}
\end{figure}

As shown in Figure \ref{fig-increase-resolution-centerline-vel}, the maximum centerline velocity loss is 14.67\% for $N_f=20$, increasing to 16.73\% and 17.47\% for $N_f=40$ and $N_f=80$, respectively.
Increasing the resolution may not mitigate the maximum velocity loss, instead, lead to a higher velocity loss.
These results prove that this velocity damping error is not a typical discretization error and can not be addressed by spatial refinement.

%

\subsubsection{With the reverse kernel gradient correction technique}

\begin{figure}[htb!]
	\centering
	\includegraphics[trim = 4.4cm 0cm 0cm 5.8cm, clip,width=1.0\textwidth]{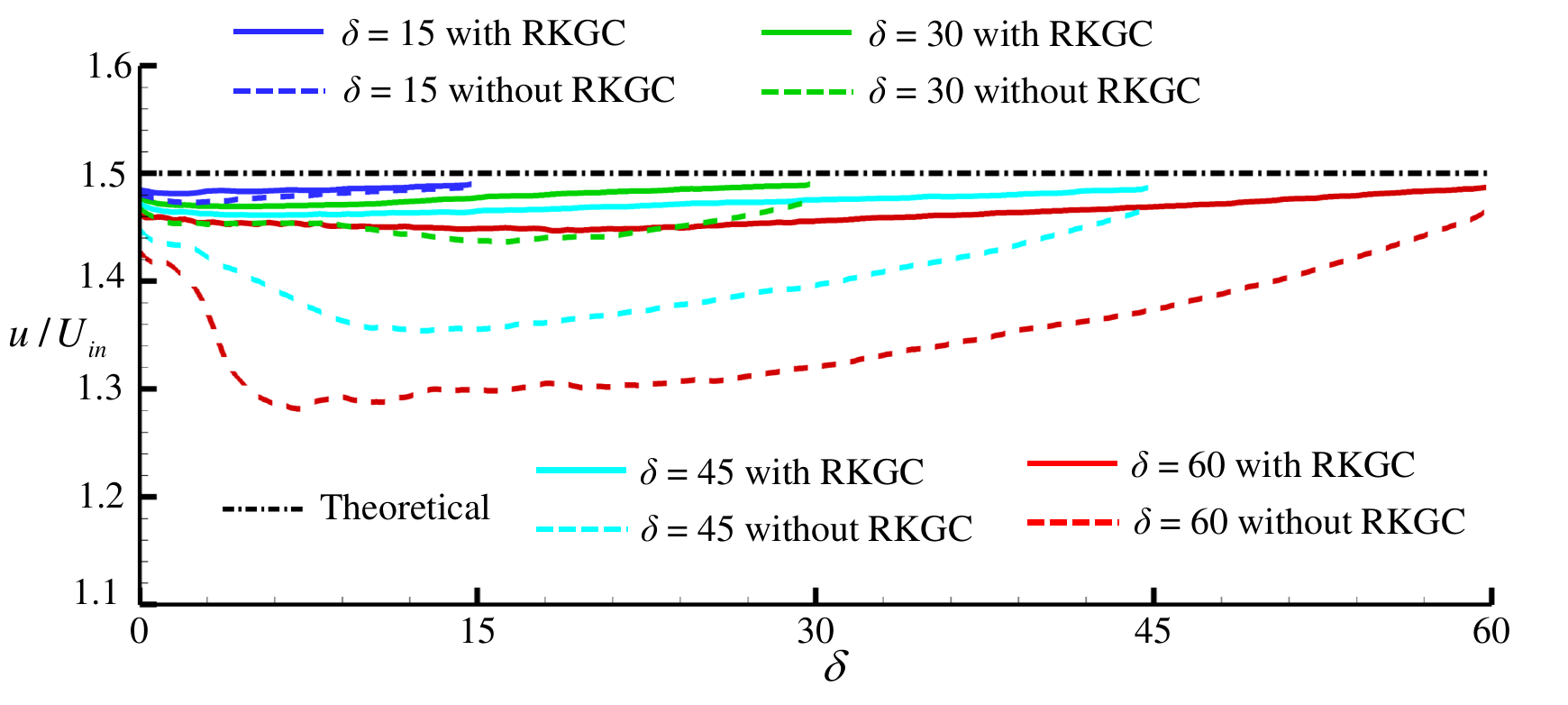}
	\caption
	{
		Laminar channel flow:
		the centerline velocity profiles of the laminar channels of different lengths($\delta=60, 45, 30, 15$) calculated by the WCSPH method with or without RKGC.
	}
	\label{fig-centerline-vel-with-without-RKGC}
\end{figure}

Since neither reducing the outlet pressure nor increasing the spatial resolution effectively mitigates the residue-induced velocity loss in a very long channel, we finally investigate the impact of the reverse kernel gradient correction (RKGC) technique.
The laminar straight channel case, as described in Section \ref{subsubsection-Channel-length}, is tested both with and without the RKGC technique.
Figure \ref{fig-centerline-vel-with-without-RKGC} presents the centerline velocity profiles along the mainstream direction.

\begin{figure}[htb!]
	\centering
	\includegraphics[trim = 0.3cm 0cm 0.3cm 1.5cm, clip,width=1.0\textwidth]{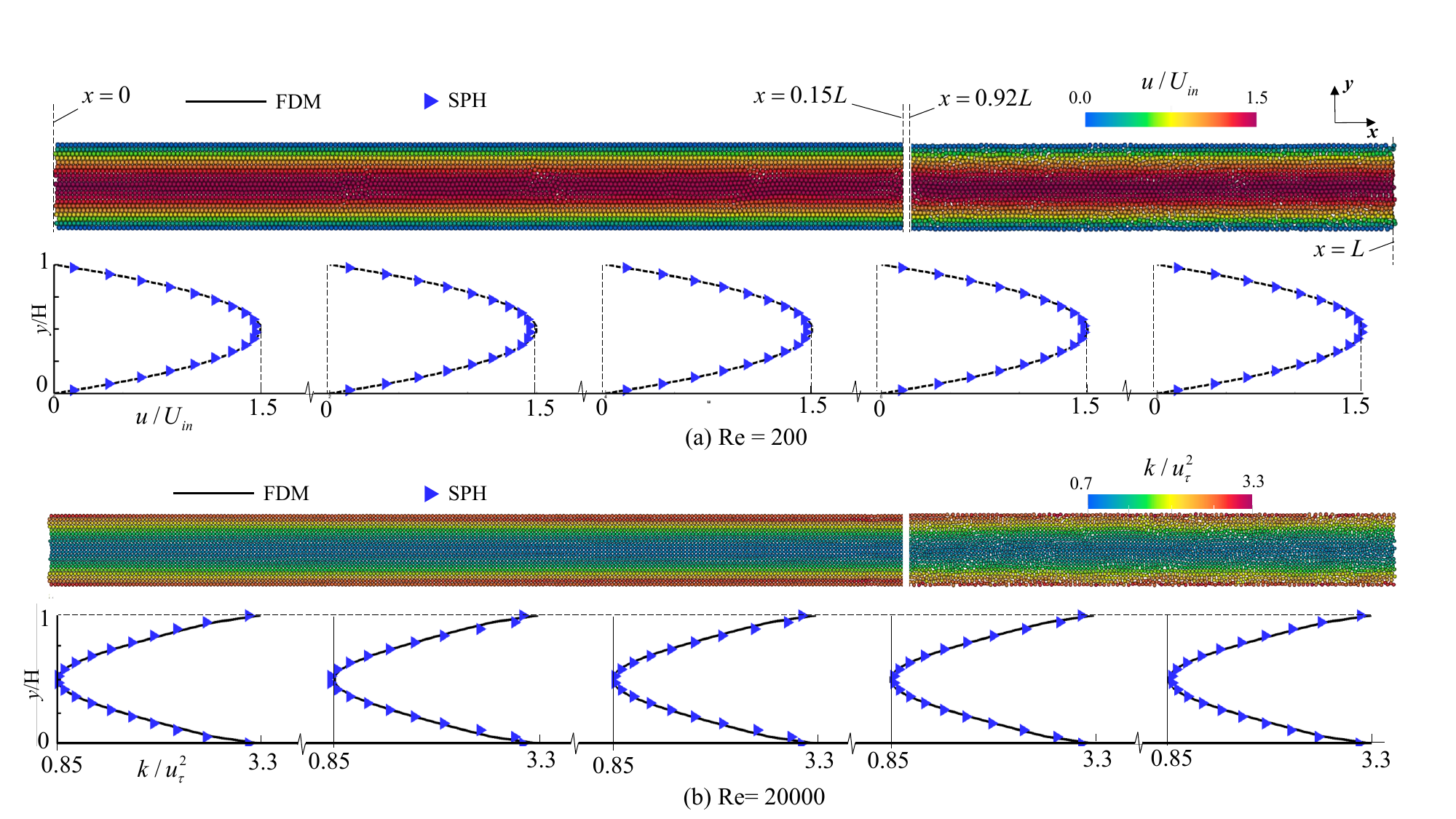}
	\caption
	{
		The results of a very long channel ($\delta=60$) calculated by the WCSPH method with the RKGC at the laminar and turbulent situations,
		and the cross-sectional profiles are obtained at positions $x/L=$ 0, 0.25, 0.5, 0.75.
		(a) The laminar case regarding the velocity;
		(b) the turbulent case regarding the turbulent kinetic energy.
	}
	\label{fig-with-RKGC-vel-k-combine}
\end{figure}

The results clearly show that the maximum centerline velocity loss is significantly reduced with the RKGC scheme.
The velocity contour together with the five cross-sectional profiles for the long ($\delta=60$) laminar case are presented in Fig \ref{fig-with-RKGC-vel-k-combine} (a).
Compared with the results without the correction, Fig. \ref{fig-laminar-vel-contours-different-lengths} (d), the corrected results agree well with the analytical solution.
The specific velocity loss data for the four channel lengths, with and without the RKGC scheme, are summarized in table \ref{table-C-WCSPH-RKGC-Result}.

However, it is worth noting that even with the RKGC scheme, a velocity loss of about 3.5\% remains at $\delta=60$, and this loss appears to increase further with channel length.
This means, similar as the analysis in the gravity-driven free-surface flow, the RKGC technique also has limitations in the channel flows.
Therefore, caution is needed when simulating very long channels using the conservative SPH method, and simplified approaches, such as reduced-order models (ROMs), may offer a more suitable alternative.

We also test the long ($\delta=60$) straight turbulent channel with the RKGC scheme, and the simulation settings are the same as those in Section \ref{subsubsection-Channel-length}.
The contour of the turbulent kinetic energy and the corresponding cross-sectional profiles are shown in Fig. \ref{fig-with-RKGC-vel-k-combine} (b).
The under-prediction is well mitigated, compared with the results without the correction (Fig. \ref{fig-vel-k-C-WCSPH}).

\begin{table}
	\scriptsize
	\centering
	\caption{The percentage of the centerline velocity loss (based on the theoretical value) for the laminar channels of the different lengths calculated by the WCSPH method with or without RKGC.}
	\renewcommand{\arraystretch}{1.2}
	\begin{tabular}{l c c}
		\hline
		$\delta$ & without RKGC & with RKGC \\
		\hline
		15       & 1.797        & 1.259     \\
		30       & 4.229        & 2.039     \\
		45       & 9.726        & 2.593     \\
		60       & 14.576       & 3.531     \\
		\hline
	\end{tabular}
	\label{table-C-WCSPH-RKGC-Result}
\end{table}

%

\section{The residue in the FDA nozzle}
\label{section-fda-channel}

\begin{figure}[htb!]
	\centering
	\includegraphics[trim = 0.8cm 0cm 0cm 7.18cm, clip,width=1.0\textwidth]{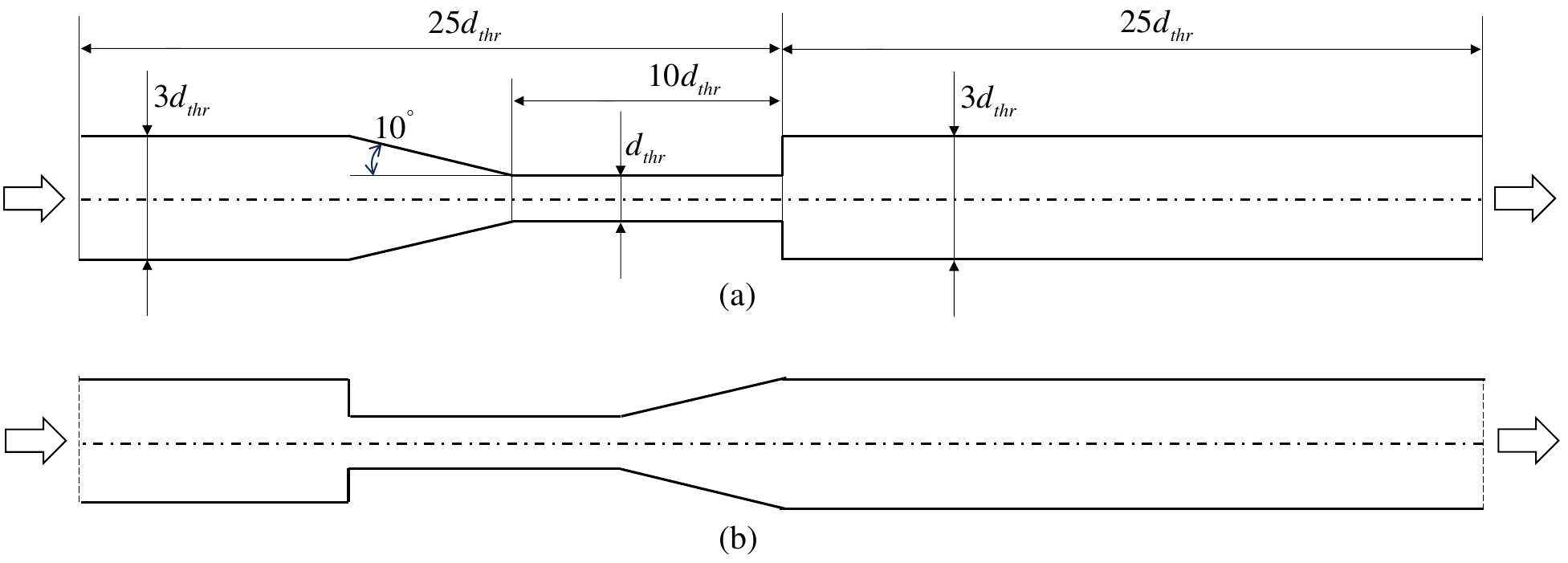}
	\caption
	{
		The geometry of the (a) FDA nozzle (b) reversed FDA nozzle.
	}
	\label{fig-fda-geo}
\end{figure}

To reveal the significance of the zero-order residue issue in a broader engineering context, we test another widely-used benchmark case, named as the FDA nozzle, to further study the influence of the residue.
The FDA nozzle case, proposed by the U.S. Food and Drug Administration(FDA), is commonly used to validate numerical solvers in reproducing flow characteristics in a complex geometry.
The geometry, as depicted in Fig. \ref{fig-fda-geo}, consists of a contraction followed by an abrupt expansion.
Extension of the inlet and outlet regions is usually necessary to exclude the boundary influence.
As a result, the total length of the channel may greatly exceed the span-wise dimension, which indicates that this case may also suffer from the zero-order residue problem.
Note that due to the lack of the Lagrangian turbulence model which can handle the strong backflow in this case, we merely simulate the laminar situations.

First, we conduct the 2D simulation to study the influence of the complex geometry on the residue, and the two geometries are tested, (1) the FDA nozzle shape; (2) the reversed FDA nozzle shape (as shown in Fig. \ref{fig-fda-geo} a and b).
The reversed FDA nozzle case is designed to comparatively study the influence of the sequence of gradual and sudden contractions.
The number of the fluid particles across the inlet channel height is 60, and the Reynolds number is 30, based on the inlet channel height and inflow mean bulk velocity, in this 2D simulation.

\begin{figure}[htb!]
	\centering
	\includegraphics[trim = 4.4cm 0cm 0cm 6.1cm, clip,width=1.0\textwidth]{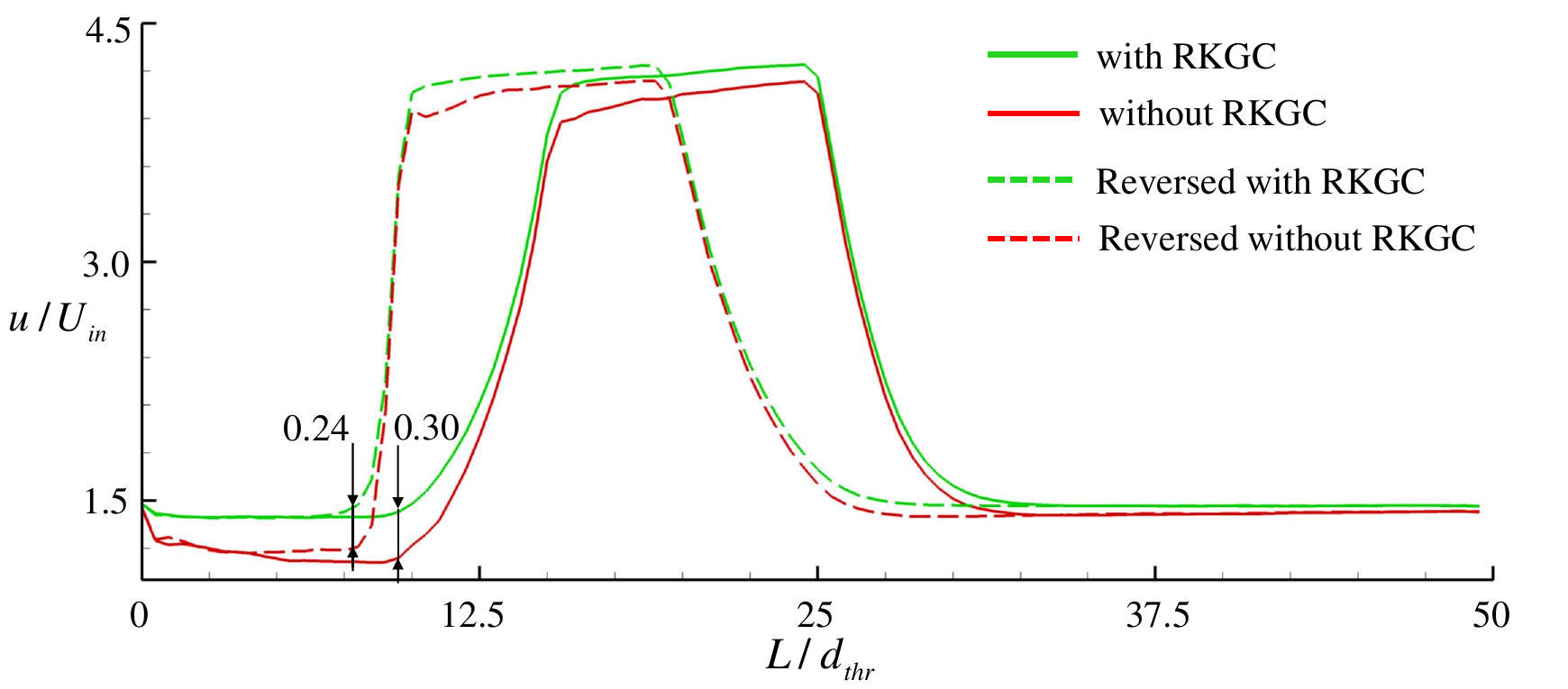}
	\caption
	{
		Two-dimensional FDA nozzle flow:
		the axis velocity profiles of the 2D FDA and reversed FDA nozzle cases calculated by the WCSPH method with or without the RKGC, the maximum velocity losses are denoted.
	}
	\label{fig-fda-2D-vel}
\end{figure}

Figure \ref{fig-fda-2D-vel} shows the centerline velocity profiles calculated by the WCSPH method with and without the RKGC technique.
For each geometry, the maximum centerline velocity differences are also presented in the figure.
The geometric change near the inlet contributes most to the residue-induced velocity damping, followed by the contraction segment.
Moreover, in the expansion segment, the velocity losses for the two shapes become relatively negligible.

\begin{figure}[htb!]
	\centering
	\includegraphics[trim = 4.4cm 0cm 0cm 6.1cm, clip,width=1.0\textwidth]{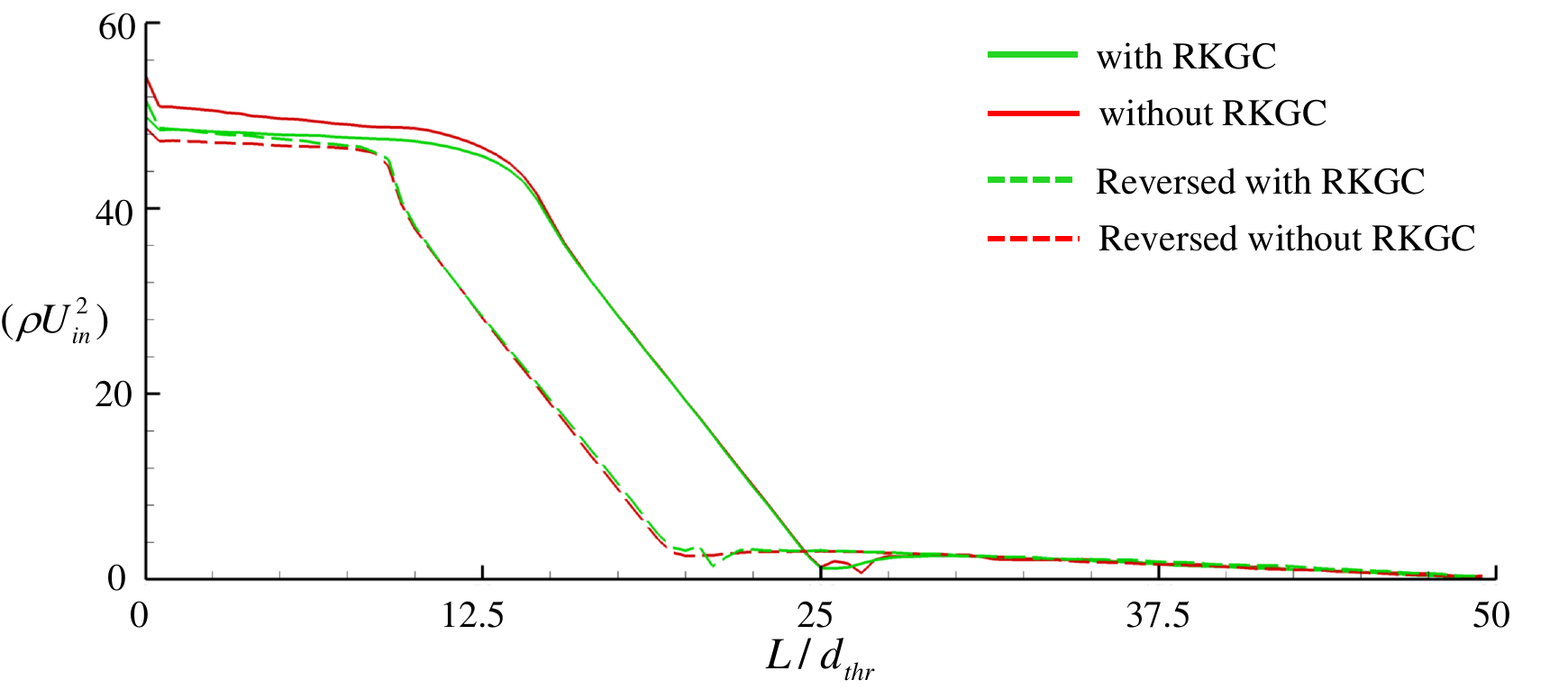}
	\caption
	{
		Two-dimensional FDA nozzle flow:
		the axis pressure profiles of the 2D FDA and reversed FDA nozzle cases calculated by the WCSPH method with or without the RKGC.
	}
	\label{fig-fda-2D-pressure}
\end{figure}

Compared with the reversed FDA nozzle case, the gradual contraction geometry in the original FDA nozzle leads to a stronger residue effect in the high background pressure region, resulting in a velocity loss up to 0.30 compared to 0.24 caused by the sudden expansion in the reversed one.
This may be attributed to the longer duration of the high local pressure, as shown in Fig. \ref{fig-fda-2D-pressure}, suggesting that a more prolonged region of high background pressure can intensify the residue-induced numerical dissipation.


\begin{figure}[htb!]
	\centering
	\includegraphics[trim = 9.4cm 0cm 0cm 2.1cm, clip,width=1.0\textwidth]{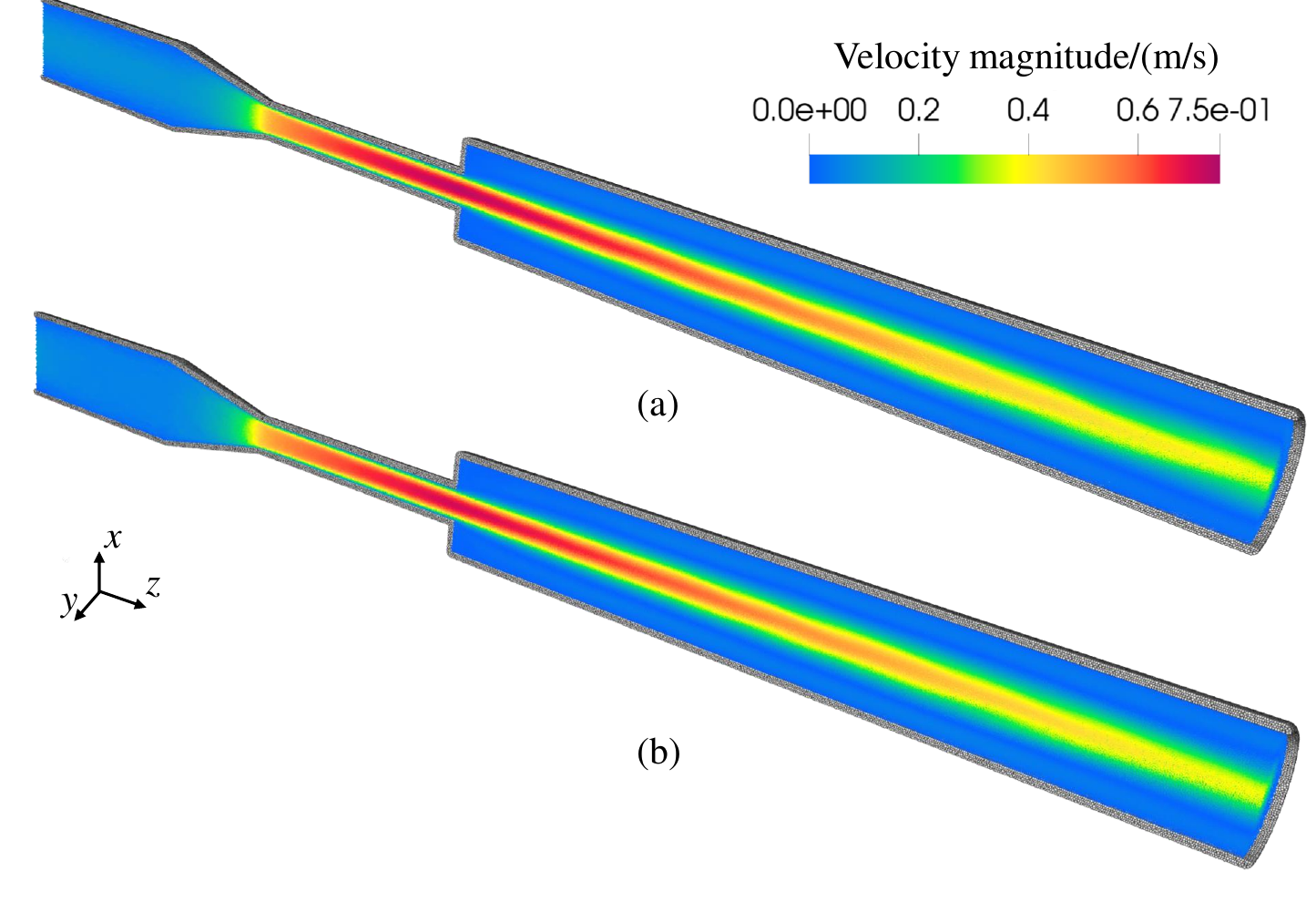}
	\caption
	{
		Three-dimensional FDA nozzle flow:
		the velocity contours sliced at the $x$-plane calculated by the WCSPH method (a) with RKGC (b) without RKGC.
	}
	\label{fig-fda-3D-vel-contour}
\end{figure}

\begin{figure}[htb!]
	\centering
	\includegraphics[trim = 4.1cm 0cm 0cm 6cm, clip,width=1.0\textwidth]{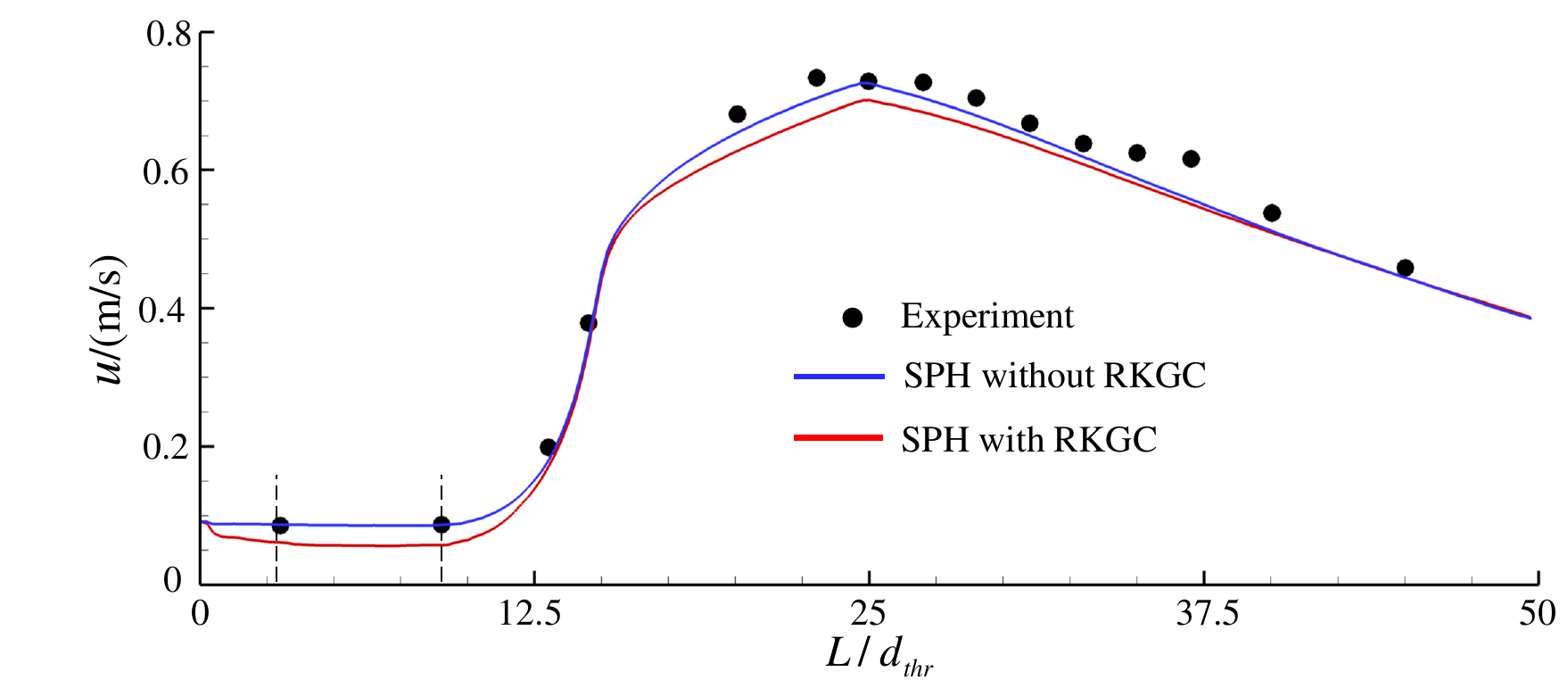}
	\caption
	{
		Three-dimensional FDA nozzle flow:
		the comparison of the axis velocity profiles calculated by the WCSPH method with RKGC or without RKGC, and the corresponding experiment data.
		The dotted lines present the monitoring positions for the cross-sectional data.
	}
	\label{fig-fda-3D-vel-axis}
\end{figure}

Next, we simulate the full-size 3D FDA nozzle, and the flow parameters are the same as Ref. \cite{huang2022simulation} where $Re=500$, the same as the experiment\cite{hariharan2011multilaboratory}.
The WCSPH method with or without the RKGC are adopted to conduct the simulations, respectively, and the particle diameter is $dp=0.0002$m.
The cross-sectional velocity contours are shown in Fig. \ref{fig-fda-3D-vel-contour} and the axis velocity profiles along the flow direction($z$) are presented in Fig. \ref{fig-fda-3D-vel-axis}.
Without the gradient correction technique, the max axis velocity loss could reach 33.7\% compared with the experiment results, while this number decreases to 2.1\% with the use of the RKGC.

\begin{figure}[htb!]
	\centering
	\includegraphics[trim = 0.0cm 0cm 0cm 0.0cm, clip,width=1.0\textwidth]{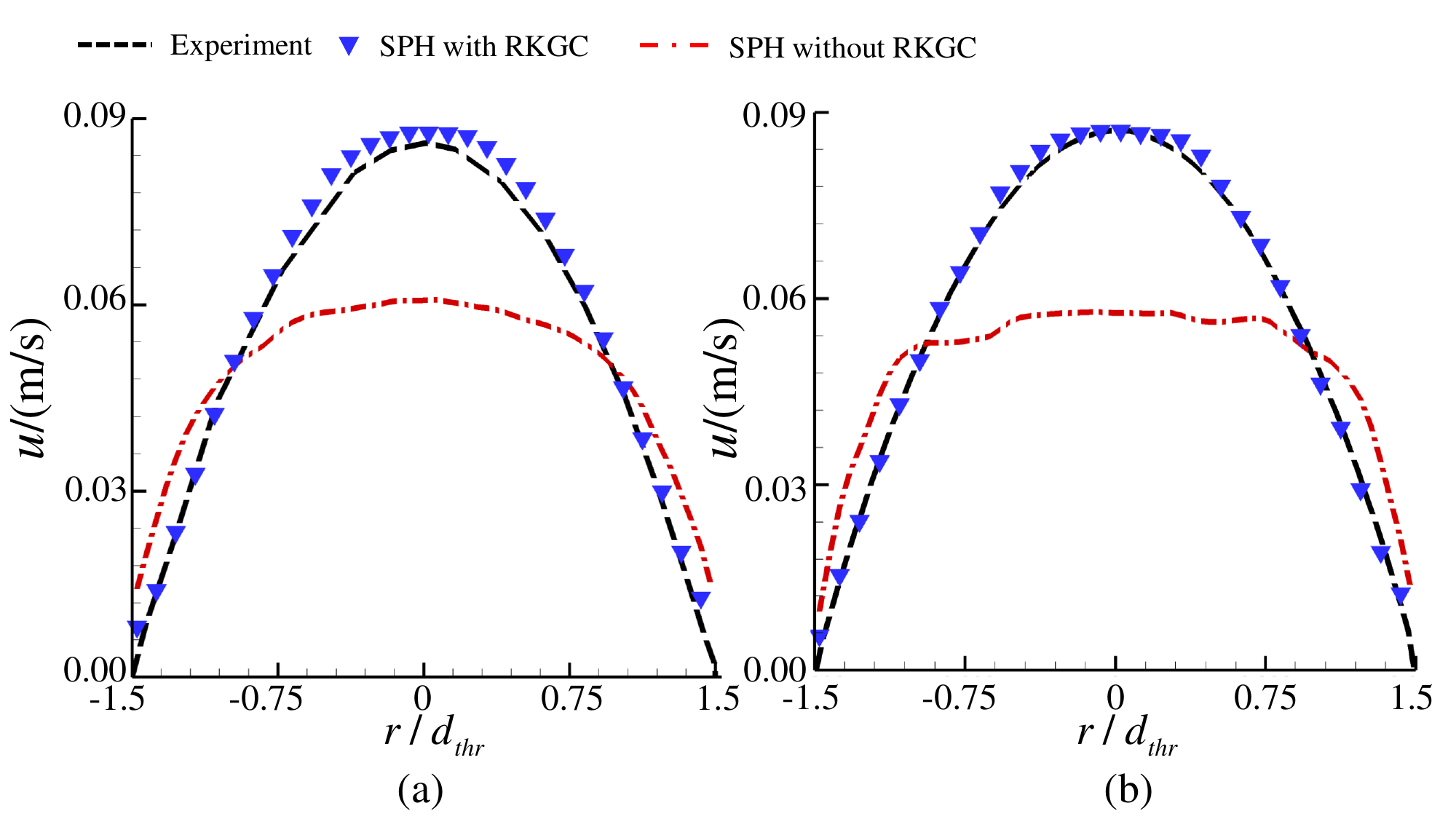}
	\caption
	{
	Three-dimensional FDA nozzle flow:
	the comparison of the cross-sectional velocity profiles, at $L/d_{thr}=$ $(a)$ 3 $(b)$ 9, calculated by the WCSPH method with RKGC or without RKGC, and the corresponding experiment data.
	}
	\label{fig-fda-3D-vel-profiles}
\end{figure}

We also extract the cross-sectional velocity profiles (along $y$ axis) at the two typical positions, as shown in Fig. \ref{fig-fda-3D-vel-profiles}.
The velocity profile is largely damped before the contraction parts without the correction.
While after using the gradient correction scheme, the velocity profiles agree well with the references.
The above findings highlight that the zero-order residue severely compromises the accuracy in this representative engineering case, underscoring the necessity of employing the correction scheme, particular in cases where the high background pressure is inevitable.
%


\section{Conclusion}
\label{section-conclusion}
This study investigates the long-standing zero-order consistency issue in the conservative smoothed particle hydrodynamics (SPH), emphasizing its detrimental effects on pressure-driven channel flows and gravity-driven free-surface flows.
Through a combination of theoretical analysis and systematic numerical experiments, we identify the residue of zero-order gradient inconsistency as the root cause of non-physical numerical damping in both scenarios.
Notably, the background pressure is found to exacerbate this residue, leading to excessive numerical dissipation and significant flow attenuation.

In the case of gravity-driven free-surface flows, our tests based on the standing wave case show that increasing the water depth or the initial wave amplitude amplifies the residue effect, resulting in the non-physical energy loss.
Besides, the laws of the energy decay caused by the two factors are concluded.
The non-dimensionless energy dissipation rate caused by the increase of the water depth exhibits remarkable similarity, which is well explained by dimensional analysis.
For pressure-driven channel flows, such as laminar and turbulent channel flows, the loss of velocity is demonstrated to be sensitive to the domain lengths and imposed outlet pressure, while insensitive to the resolution.
Adjusting the outlet pressure is proved to be effective in reducing the velocity loss, but this remedy may cause numerical instability, and hence is not feasible.

To address zero-order residue issue in the high background situations, such as the flows in a deep water tank or a long channel, we implement the reverse kernel gradient correction(RKGC) technique.
The RKGC scheme shows effectiveness in mitigating the residue-induced damping across both the flow types.
However, our findings highlight that while this correction approach reduces the adverse effects to a degree, it is fundamentally constrained by its limited ability to fully restore gradient consistency when the background pressure is too large.
Therefore, in scenarios involving substantial background pressure, including the deep water or the long channel flows, the conservative SPH method should be applied with caution, and simplified approaches such as potential flow theory or reduced-order models may serve as more effective and physically consistent alternatives.

Finally, the FDA nozzle benchmark case demonstrates that in engineering applications with inherently high background pressure and intricate geometries, the residue effect becomes pronounced and cannot be neglected, further proving the necessity of using the correction scheme.

%
%
\section*{Appendix A. The values of the coefficients}
\label{appendix}
\begin{table}
	\scriptsize
	\centering
	\caption{Coefficients for the WCSPH METHOD.}
	\begin{tabularx}{8.5cm}{@{\extracolsep{\fill}}lc}
		\hline
		Name of the coefficients & Value  \\
		\hline
		$C_1$                    & $1.44$ \\
		\hline
		$C_2$                    & $1.92$ \\
		\hline
		$C_\mu$                  & $0.09$ \\
		\hline
		$\sigma_k$               & $1.0$  \\
		\hline
		$\sigma_\epsilon$        & $1.3$  \\
		\hline
	\end{tabularx}
	\label{tab-coefficients}
\end{table}

\bibliographystyle{elsarticle-num}
\bibliography{reference}
%
%
\end{document}